\documentstyle[graphicx,times,amssymb]{mn}

\begin{document}

\title{X-ray colour maps of the cores of galaxy clusters}
\author[J.S. Sanders, A.C. Fabian and S.W. Allen] {J.S. Sanders,
  A.C. Fabian and S.W. Allen
  \\
  Institute of Astronomy, Madingley Road, Cambridge. CB3 0HA}
\maketitle

\begin{abstract}
We present an analysis of X-ray colour maps of the cores of clusters
of galaxies, formed from the ratios of counts in different X-ray
bands. Our technique groups pixels lying between contours in an
adaptively-smoothed image of a cluster. We select the contour levels
to minimize the uncertainties in the colour ratios, whilst preserving
the structure of the object.  We extend the work of Allen \& Fabian
(1997) by investigating the spatial distributions of cooling gas and
absorbing material in cluster cores. Their sample is almost doubled:
we analyse archive \emph{ROSAT} PSPC data for 33 clusters from the
sample of the 55 brightest X-ray clusters in the sky. Many of our
clusters contain strong cooling flows. We present colour maps of a
sample of the clusters, in addition to adaptively-smoothed images in
different bands. Most of the cooling flow clusters display little
substructure, unlike several of the non-cooling-flow clusters.

We fitted an isothermal plasma model with galactic absorption and
constant metallicity to the mid-over-high energy colours in our
clusters. Those clusters with known strong cooling flows have inner
contours which fit a significantly lower temperature than the outer
contours. Clusters in the sample without strong cooling flows show no
significant temperature variation. The inclusion of a metallicity
gradient alone was not sufficient to explain the observations. A
cooling flow component plus a constant temperature phase did account
for the colour profiles in clusters with known strong cooling flow
components.  We also had to increase the levels of absorbing material
to fit the low-over-high colours at the cluster centres. Our results
provide more evidence that cooling flows accumulate absorbing
material. No evidence for increased absorption was found for the
non-cooling-flow clusters.
\end{abstract}


\section{Introduction}
Clusters of galaxies are among the largest dynamical objects in the
universe. They contain large quantities of virilised gas (the
intracluster medium, or ICM) at temperatures of the order of
$10^7$--$10^8$~K, heated by their initial gravitational collapse. Due
to this gas they are powerful emitters of X-rays from bremsstrahlung
and line emission. The gas masses of rich clusters exceed $10^{14} \:
M_{\odot}$, and they can emit over $10^{45} \textrm{ erg s}^{-1}$ in
X-rays.

The radiative cooling time, $t_{\mathrm{cool}}$, is shortest
in the densest regions, at the centre of a cluster, and is often found
to be less than the plausible age of the
cluster, i.e. the time since its last major interaction, which is a
significant fraction of the Hubble time. As the gas cools, its
temperature decreases, but its density must rise to maintain
hydrostatic equilibrium. This is achieved by the gas above flowing in.
This is known as a \emph{cooling flow} (See Fabian 1994 for a review).

Observations have shown that cooling flows are
\emph{inhomogeneous}. The mass drops out of the flow into a form which
does not emit X-rays. The exact destination and final
form of this gas has not been determined, but magnetic fields are
probably important in determining, and constraining, the structure of
cooled gas.

The rate of the mass of gas cooling out at a certain radius is known as
the mass deposition rate, $\dot{M}(r)$. This value can total from
10--1000 $M_{\odot} \textrm { yr}^{-1}$. The mass deposition rate can
be estimated by assuming that all the luminosity within the cooling
radius, where the cooling time is less than the age of the cluster, is
due to the cooling gas radiation, plus the $P \mathrm{d}V$ work done
on the gas as it enters that radius. The relationship between the mass
deposition rate and radius is found to be roughly linear, as the
surface brightness profiles are not as centrally peaked as would be
expected for a homogeneous cooling flow.

Allen \& Fabian (1997) undertook a study of the spatial distributions
of cooling gas and intrinsic X-ray absorbing material in a sample of
nearby clusters, using X-ray colour profiles, formed from the ratio of
flux in selected bands, using data from the \emph{ROSAT} Position
Sensitive Proportional Counter (PSPC).
They found that the profiles indicated that there were significant
central concentrations of cooling gas in their cooling flow clusters,
becoming approximately isothermal at larger radii. Their profiles also
showed large levels of X-ray absorption in the cluster cores, increasing
with decreasing radius.

The bands they examined are listed in Table \ref{tab:bands}, which
they combined to form two X-ray colour ratios, $B/D$ and $C/D$. To
analyse the observations, they calculated theoretical curves for the
colours as functions of temperature and metallicity, with a uniform
screen to model the absorbing column density. They modelled both
colours with isothermal and cooling gas models. In Section
\ref{sec:models} we reevaluate these functions using more recent
plasma models.

To demonstrate the presence of distributed cooling gas in the central
regions of cooling flow clusters, they used the $C/D$ profile, and the
$B/D$ ratio to demonstrate the presence of intrinsic X-ray absorbing
gas. They found the absorbing material only partially covers the
X-ray emitting region.

\begin{table}
  \caption{Definition of the energy bands (Allen \& Fabian 1997)}
  \begin{tabular}{ll}
    Band & Energy range (keV) \\ \hline
    $A_1$ & 0.10--0.20 \\
    $A_2$ & 0.20--0.40 \\
    $B$   & 0.41--0.79 \\
    $C$   & 0.80--1.39 \\
    $D$   & 1.40--2.00 \\
    $F$   & 0.41--2.00 \\ \hline
  \end{tabular}
  \label{tab:bands}
\end{table}

In this paper, we extend the analysis of Allen \& Fabian
(1997). Instead of making a simple radial profile of each cluster, we
have developed a technique to take account of the two-dimensional
structure of the image. Simple binning works in areas of high flux,
where the errors created in forming ratios of counts are small. In low
flux regions, the number of pixels required to bin destroys most of
the spatial information in the image. Our method attempts to make
statistically useful X-ray colours, whilst preserving spatial resolution.

\section{Observations and colour map analysis}
\label{sec:obs}
We analysed archival \emph{ROSAT} PSPC data for a sample of 33
clusters. The clusters we
examined are shown in Table \ref{tab:clusters}.  The sample was mainly
chosen to include known strong cooling flow clusters (from Peres et
al. 1998), with a few non-cooling-flow clusters for comparison.

We used the bands of Allen \& Fabian (1997) in our analysis.
These bands are optimised for the study of cooling gas. This
precluded the use of the extended object analysis package of Snowden
\& Kuntz (1998), although we made use of some modified versions of
a few of their routines.

The full analysis procedure for each cluster is as follows.
\begin{enumerate}
\item We removed time periods from the observation where the
  Master Veto count rate averaged over $\sim 30$ s was greater than 220
  counts $\mathrm{s}^{-1}$ (Snowden et al. 1994).
\item A modified version of Snowden's \textsc{cast\_data}
  routine was used to create $512 \times 512$ pixel images of the cluster, at a
  pixel size of approximately $15 \times 15 \textrm{ arcsec}^2$. We created
  images of the cluster in the $B$, $C$ and $D$ bands from Table
  \ref{tab:bands}.
\item We corrected the $B$, $C$ and $D$ band images for background
  counts (Section \ref{sec:background}).
\item An \emph{adaptively-smoothed} image (Section
  \ref{sec:adapt}) was made of the central region of the cluster, in the high
  energy $D$ band. We created an image of size $64 \times 64$, or
  occasionally $128 \times 128$ pixels, depending on the extent of the
  cluster.
\item We created linear contours in the adaptively-smoothed image, by
  simply taking the difference between the maximum and minimum count
  rates, and dividing by the number of contours.
\item The counts for those pixels between each two
  neighbouring contours was added, in each of the background corrected $B$, $C$
  and $D$ images.
\item We used the count totals to generate average $B/D$ and $C/D$
  colours for each contour. The error on each result is:
  \begin{eqnarray}
    \label{eq:errorcontour}
    \left( \frac{\sigma_r}{r} \right)^2
    & = &
    \frac{ \sum_{\mathrm{pixels}} \left( N_i^X + C_i^X \right) }
         { \left[ \sum_{\mathrm{pixels}} \left( N_i^X - C_i^X \right)
           \right]^2 }
         + \nonumber \\ & &
    \frac{ \sum_{\mathrm{pixels}} \left( N_i^Y + C_i^Y \right) }
         { \left[ \sum_{\mathrm{pixels}} \left( N_i^Y - C_i^Y \right)
           \right]^2 },
  \end{eqnarray}
  where $r$ is the $X/Y$ average ratio, and $\sigma_r$ is the error on
  the ratio. $N_i^X$ is the uncorrected number of photons in pixel $i$
  in band $X$, and $C_i^X$ is the background correction for pixel $i$
  in band $X$. This expression does not take account of systematic
  background uncertainties. We ignored pixels that were suspected to
  contain point sources (Section \ref{sec:background}) outside the
  core.
\item A Monte Carlo procedure was used to change the contour levels in
  the adaptively-smoothed image to minimize the total error
  squared in the contours for both colours, or the value
  \begin{equation}
    x = \sum_{\mathrm{contours}} \left\{
      \left[ \sigma_{(B/D)_i} \right] ^2 + 
      \left[ \sigma_{(C/D)_i} \right] ^2
      \right\}.
  \end{equation}
  The result of moving the contour levels to minimize $x$ creates contour
  levels which contain similar numbers of counts and signal to
  noise ratios. The procedure produces colour ratios with roughly equal
  fractional uncertainties for each contour.
\end{enumerate}

We used six contours to process the cluster images.
This was a good compromise between showing detail and
keeping statistical errors low. The results were stable to small changes
in the number of contours.

\subsection{Adaptive smoothing}
\label{sec:adapt}
Adaptive smoothing is a technique to smooth an image and retain
significant small-scale structure (Ebeling, White \& Rangarajan
2000). Smoothing is required because it is often the case, especially
in X-ray data, that there are few photons per pixel. However,
convolving an image with a Gaussian kernel leads to broadening of
features and loss of small-scale structure. An algorithm is required
which retains information on all scales.

An adaptive kernel smoothing (AKS) algorithm applies a kernel with a
varying size to an image. A part of the image with few counts is
smoothed with a large kernel, whilst a part with many counts is
smoothed with a small kernel. The algorithm decides what is the
`natural' smoothing scale for a part of an image. AKS algorithms
differ on how they choose an appropriate kernel size. We used the
\textsc{asmooth} algorithm of Ebeling, White \& Rangarajan (2000) to
adaptively smooth our cluster images. \textsc{asmooth} is unusual in
that it uses the local signal-to-noise ratio to adjust the size of the
kernel. The minimum signal-to-noise ratio of the part of the image
smoothed by the kernel is a parameter to the algorithm. A level of
6-$\sigma$ seemed to produce the best results; it appeared not to
generate false detail from spurious pixels. We used a Gaussian rather
than top-hat adaptive kernel.

Other methods for revealing structure in data include wavelet analysis
(Starck \& Pierre 1998). Different methods have other advantages, but
the \textsc{asmooth} algorithm is particularly simple and quick to
perform.

\subsection{Background removal}
\label{sec:background}
The raw data contains counts not from the observed source. These
background counts, sources of which are listed below, must be removed.
\begin{enumerate}
\item The X-ray background, which
  comes through the aperture of the telescope, and is
  observed with the same vignetting response of the telescope as the
  object. This includes solar and extragalactic background emission.
\item The particle background (Plucinsky et al. 1993) is diffuse and
  is caused by high energy cosmic rays. The flux of the particle
  background is approximately constant across the detector.
\end{enumerate}
These components vary in strength with X-ray energy, so they must be
removed to make X-ray colours. 

To estimate the background we took the average count from the 30--35
arcmin region, beyond the `ribs' in the field of the PSPC. From this
count we subtracted an estimate of the particle background from
Plucinsky et al. (1993), of $4 \times 10^{-6}$ counts
$\textrm{pixel}^{-1}$ $\textrm{s}^{-1}$ $\textrm{arcmin}^{-2}$
$\textrm{keV}^{-1}$. We corrected the result for vignetting.
The corrected count and the particle background was removed from all
pixels.

Accounting for the total background reduces the scatter of the
observed data points. It also hardens the emission from the outermost
parts of clusters which emit few photons.

Three clusters, A2142, A1795 and A2029, contained obvious point
sources in their fields outside the cluster core.  We removed the
point sources using a mask which excluded the affected region around
them. Those pixels inside the mask were not included in the sum of
counts inside a contour.

\begin{table*}
  \begin{minipage}{140mm}
    \caption{The sample of clusters examined. $z$ is the redshift taken
      from the NASA/IPAC
      Extragalactic Database (NED), and the values of $\dot{M}$, the
      mass deposition rate, are
      PSPC observation derived values taken from Peres et
      al. (1998), as are the central cooling time values,
      $t_{\mathrm{cool}}$. 
      The sequence identification and exposure time give
      PSPC observation details. $S(B/D)$ and $S(C/D)$ are the significance
      of a colour gradient, larger
      values imply a larger colour gradient (Section \ref{sec:results}). The
      clusters are listed in cooling flow flux order, $\dot{M}/z^2$.}
    \begin{tabular}{llrlcccc}
      Cluster     & Sequence ID & Exposure/s & $z$ &  $\dot{M}/(M_{\odot} \: \mathrm{yr}^{-1})$ &
        $t_{\mathrm{cool}}/ \textrm{Gyr}$ &
      $S(B/D)$& $S(C/D)$ \\ \hline
      Virgo       & rp800187n00 & 10536   & 0.0037 & $39^{+2}_{-9}$      & $0.2^{+0.0}_{-0.0}$   &  4.9 &  21  \\
      Perseus     & rp800186n00 & 4711    & 0.0183 & $556^{+33}_{-24}$   & $0.9^{+0.0}_{-0.0}$   &  5.9 &  8.4 \\
      2A 0335+096 & rp800083n00 & 10220   & 0.0349 & $325^{+32}_{-43}$   & $0.9^{+0.0}_{-0.0}$   &  2.6 &  7.1 \\
      Centaurus   & rp800192n00 & 7793    & 0.0109 & $30^{+10}_{-5}$     & $0.8^{+0.0}_{-0.0}$   &  1.9 &  14  \\
      A2199       & rp800644n00 & 40999   & 0.030  & $154^{+18}_{-8}$    & $1.9^{+0.0}_{-0.1}$   &  7.8 &  10  \\
      Ophiuchus   & rp800279n00 & 3932    & 0.028  & $127^{+48}_{-94}$   & $3.0^{+0.4}_{-0.3}$   & -1.2 &  0.3 \\
      Klemola 44  & rp800354n00 & 3353    & 0.0283 & $87^{+25}_{-19}$    & $2.3^{+0.6}_{-0.2}$   & -1.6 &  0.1 \\
      A2052       & rp800275n00 & 6211    & 0.0348 & $125^{+26}_{-6}$    & $2.5^{+0.1}_{-0.1}$   &  2.1 &  4.6 \\
      A262        & rp800254n00 & 8686    & 0.0164 & $27^{+4}_{-3}$      & $1.5^{+0.1}_{-0.1}$   &  2.2 &  5.1 \\
      PKS 0745-191& rp800623n00 & 10473   & 0.1028 & $1038^{+116}_{-68}$ & $2.2^{+0.1}_{-0.1}$   &  2.6 &  2.7 \\
      A1060       & rp800200n00 & 15764   & 0.0124 & $15^{+5}_{-7}$      & $4.7^{+0.4}_{-0.3}$   &  0.3 &  0.1 \\
      A1795       & rp800105n00 & 36273   & 0.0627 & $381^{+41}_{-23}$   & $1.9^{+0.1}_{-0.1}$   &  4.6 &  5.9 \\
      Hydra-A     & rp800318n00 & 18398   & 0.0522 & $264^{+81}_{-60}$   & $2.0^{+0.0}_{-0.0}$   &  3.8 &  5.5 \\
      A2029       & rp800249n00 & 12542   & 0.0767 & $556^{+44}_{-73}$   & $2.9^{+0.1}_{-0.1}$   &  4.4 &  5.2 \\
      A496        & rp800024n00 & 8857    & 0.033  & $95^{+13}_{-12}$    & $1.8^{+0.1}_{-0.1}$   &  1.4 &  3.8 \\
      MKW3        & rp800128n00 & 9984    & 0.0449 & $175^{+14}_{-46}$   & $3.0^{+0.1}_{-0.1}$   & -0.1 &  0.8 \\
      A478        & rp800193n00 & 21969   & 0.0882 & $616^{+63}_{-76}$   & $2.8^{+0.1}_{-0.1}$   & -0.8 &  2.1 \\
      A85         & rp800250n00 & 10238   & 0.0521 & $198^{+53}_{-52}$   & $2.4^{+0.1}_{-0.1}$   &  1.5 &  2.9 \\
      A3112       & rp800302n00 & 7598    & 0.0746 & $376^{+80}_{-61}$   & $1.9^{+0.1}_{-0.1}$   &  1.4 &  3.6 \\
      Cygnus-A    & rp800622n00 & 9442    & 0.057  & $244^{+26}_{-22}$   & $2.6^{+0.1}_{-0.1}$   &  0.6 &  1.6 \\
      AWM7        & rp800168n00 & 13088   & 0.0172 & $41^{+6}_{-6}$      & $1.9^{+0.2}_{-0.2}$   & -0.4 &  2.2 \\
      A4059       & rp800175n00 & 5439    & 0.0478 & $130^{+27}_{-21}$   & $3.4^{+0.6}_{-0.4}$   &  0.9 &  1.9 \\
      A3571       & rp800287n00 & 6062    & 0.0391 & $72^{+56}_{-31}$    & $5.8^{+0.9}_{-1.0}$   &  6.7 &  3.4 \\
      A2597       & rp800112n00 & 7163    & 0.0824 & $271^{+41}_{-41}$   & $2.3^{+0.1}_{-0.1}$   & -0.3 &  1.4 \\
      A644        & rp800379n00 & 10246   & 0.0704 & $189^{+106}_{-35}$  & $6.8^{+0.4}_{-0.4}$   &  2.8 &  1.4 \\
      A2204       & rp800281n00 & 5357    & 0.1523 & $852^{+127}_{-82}$  & $3.1^{+0.1}_{-0.1}$   &  2.8 &  3.4 \\
      A2142       & rp150084n00 & 7734    & 0.0899 & $350^{+66}_{-133}$  & $5.2^{+0.4}_{-0.3}$   &  1.0 &  1.0 \\
      A3558       & rp800076n00 & 29490   & 0.0478 & $40^{+39}_{-10}$    & $10.2^{+0.3}_{-0.2}$  & -0.2 &  1.2 \\
      A401        & rp800235n00 & 7457    & 0.0743 & $42^{+82}_{-42}$    & $10.6^{+4.0}_{-1.8}$  & -0.6 & -1.3 \\
      Coma        & rp800005n00 & 21140   & 0.0232 & $0^{+1}_{-0}$       & $17.7^{+6.7}_{-4.1}$  &  1.7 &  0.9 \\
      A754        & rp800550n00 & 8156    & 0.0542 & $0^{+29}_{-0}$      & $15.0^{+3.1}_{-2.2}$  &  1.6 &  2.5 \\
      A2256       & rp800162a01 & 4747    & 0.0581 & $0^{+14}_{-0}$      & $15.0^{+4.0}_{-3.6}$  &  0.2 &  0.9 \\
      A119        & rp800251n00 & 15197   & 0.044  & $0^{+2}_{-0}$       & $19.2^{+12.2}_{-8.6}$ & -0.9 & -1.8 \\
      \hline
    \end{tabular}
    \label{tab:clusters}
  \end{minipage}
\end{table*}

\begin{table}
\caption{Mean distance of the pixels in each contour from the cluster
centre. Distances are measured in pixels (15 arcsec). The contours
are numbered from 1 (inner) to 6 (outer). Clusters marked with an
asterix (*) were examined using a $128\times128$ pixel adaptively
smoothed image, rather than a $64\times64$ image.}
\begin{tabular}{lrrrrrr}
Cluster  &        1 &        2 &        3 &        4 &        5 &  6 \\
\hline
2A 0335+096 &   2.1 &      3.8 &      5.9 &      8.7 &     13.8 &     28.7 \\
A1060*   &      6.9 &     13.2 &     19.2 &     26.5 &     37.0 &     60.7 \\
A119*    &      7.5 &     14.6 &     20.8 &     27.5 &     37.0 &     60.4 \\
A1795    &      2.6 &      4.2 &      6.6 &      9.6 &     14.7 &     28.9 \\
A2029    &      2.7 &      4.3 &      6.4 &      9.5 &     13.7 &     28.5 \\
A2052    &      3.0 &      5.0 &      7.9 &     10.5 &     14.9 &     28.9 \\
A2142    &      3.1 &      5.2 &      8.3 &     12.2 &     18.3 &     29.5 \\
A2199    &      3.6 &      7.0 &     10.5 &     14.4 &     20.3 &     31.1 \\
A2204    &      2.8 &      2.9 &      3.6 &      5.5 &      8.0 &     26.9 \\
A2256*   &      7.4 &     11.9 &     16.7 &     20.8 &     25.6 &     56.1 \\
A2597    &      2.4 &      2.9 &      3.7 &      5.3 &      8.1 &     27.0 \\
A262*    &      6.3 &     11.8 &     18.1 &     24.8 &     33.4 &     59.1 \\
A3112    &      3.3 &      3.7 &      5.2 &      8.0 &     13.1 &     28.5 \\
A3558*   &      7.1 &     12.5 &     17.0 &     21.0 &     27.6 &     56.8 \\
A3571*   &      5.5 &     10.2 &     14.4 &     21.7 &     32.1 &     58.0 \\
A401     &      3.0 &      6.4 &      9.7 &     12.8 &     16.9 &     29.6 \\
A4059    &      3.3 &      5.7 &      8.5 &     11.5 &     17.5 &     30.3 \\
A478     &      2.4 &      3.6 &      5.1 &      7.6 &     11.2 &     27.8 \\
A496     &      2.8 &      5.7 &      9.4 &     13.9 &     20.1 &     31.0 \\
A644     &      3.8 &      6.1 &      8.7 &     11.8 &     16.4 &     29.0 \\
A754*    &     12.0 &     25.7 &     27.7 &     30.7 &     36.6 &     57.2 \\
A85      &      2.9 &      5.9 &     10.0 &     14.0 &     19.5 &     30.8 \\
AWM7*    &      7.9 &     14.4 &     21.1 &     27.7 &     36.5 &     60.0 \\
Centaurus* &    5.7 &     12.3 &     18.5 &     27.2 &     39.2 &     60.9 \\
Coma*    &     14.0 &     22.5 &     29.4 &     36.7 &     46.6 &     63.0 \\
Cygnus A &      4.6 &      5.3 &      6.2 &      8.8 &     16.5 &     29.5 \\
Hydra A  &      2.6 &      3.8 &      6.7 &      9.6 &     13.4 &     28.4 \\
KLEM44   &      4.3 &      6.0 &      8.9 &     12.4 &     17.8 &     29.8 \\
MKW3     &      2.4 &      4.7 &      7.0 &      9.2 &     12.9 &     28.2 \\
Ophiuchus &     7.3 &     12.2 &     15.2 &     19.1 &     24.4 &     31.3 \\
Perseus* &      4.2 &      8.4 &     13.5 &     22.2 &     34.8 &     59.3 \\
PKS 0745-191 &  2.4 &      3.1 &      4.0 &      6.2 &      9.6 &     27.3 \\
Virgo*   &      7.4 &     14.2 &     21.7 &     30.2 &     41.1 &     62.2 \\
\hline
\end{tabular}
\label{tab:disttab}
\end{table}

\section{Results}
\label{sec:results}
\begin{figure*}
  \includegraphics[angle=-90,width=0.28\textwidth]{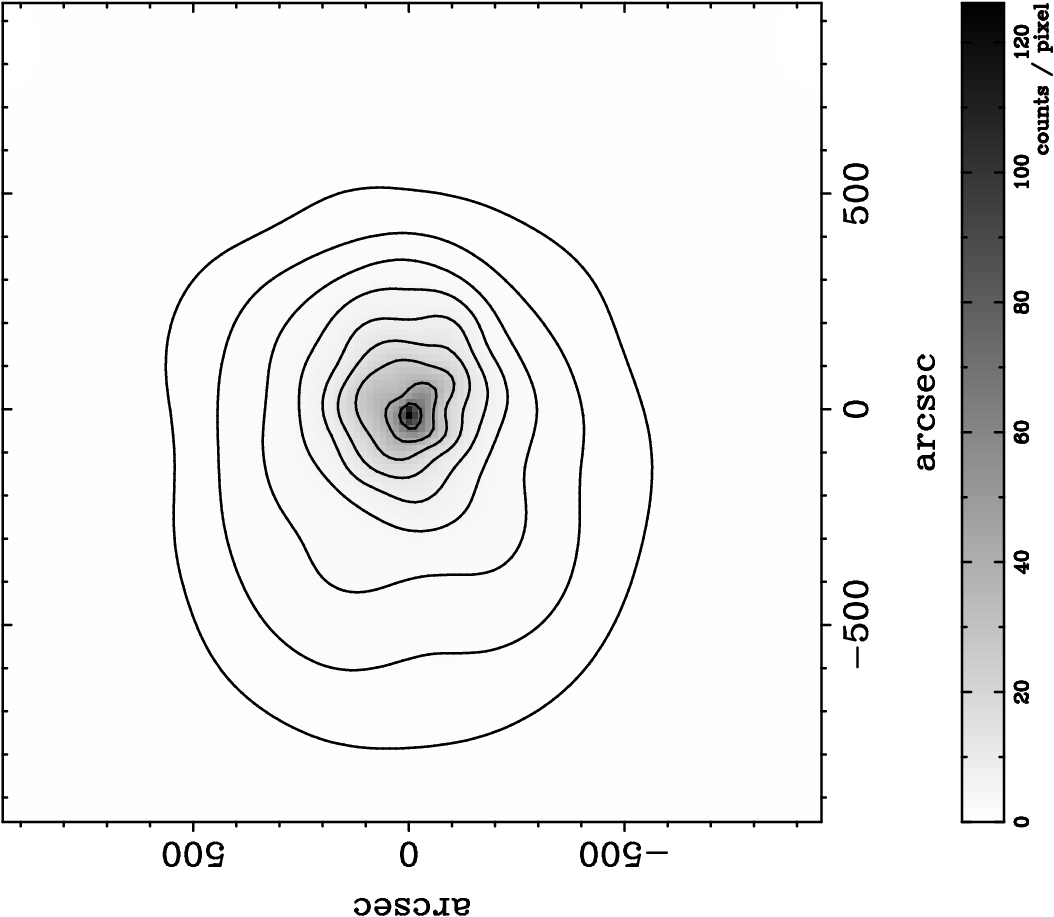}
  \hspace{6mm}
  \includegraphics[angle=-90,width=0.28\textwidth]{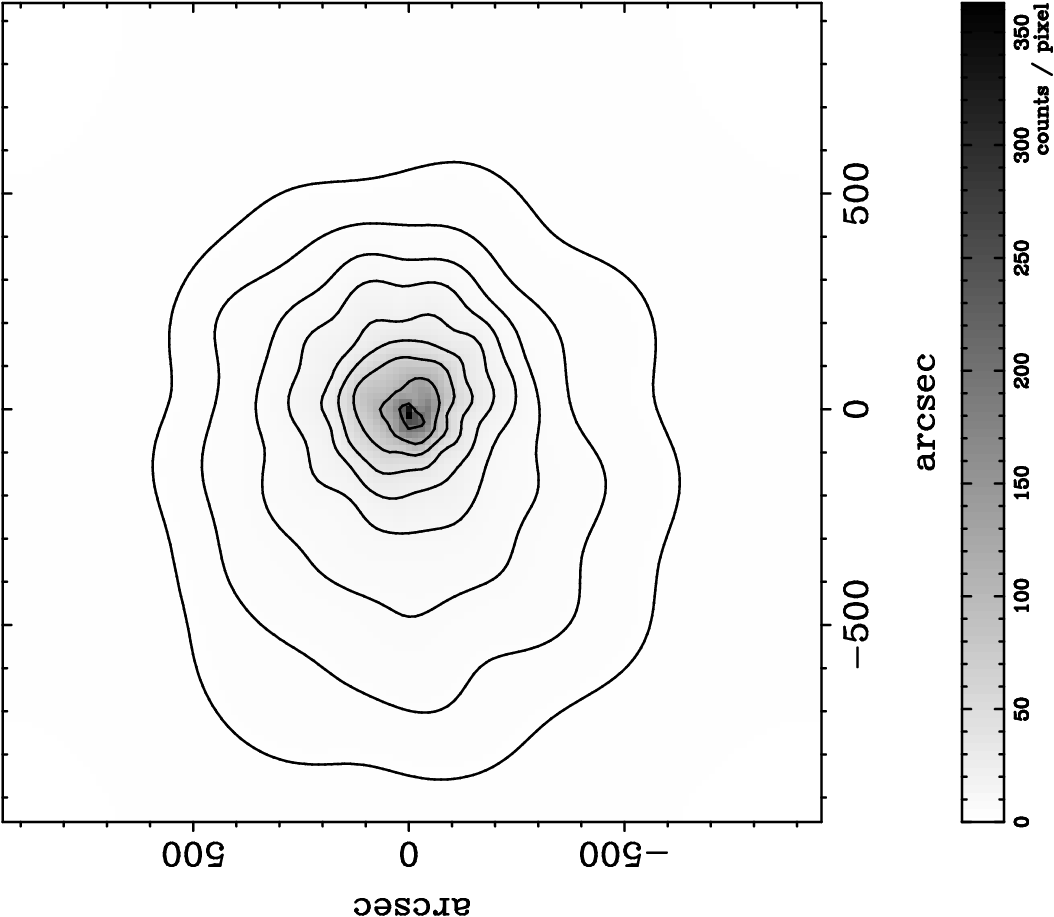}
  \hspace{6mm}
  \includegraphics[angle=-90,width=0.28\textwidth]{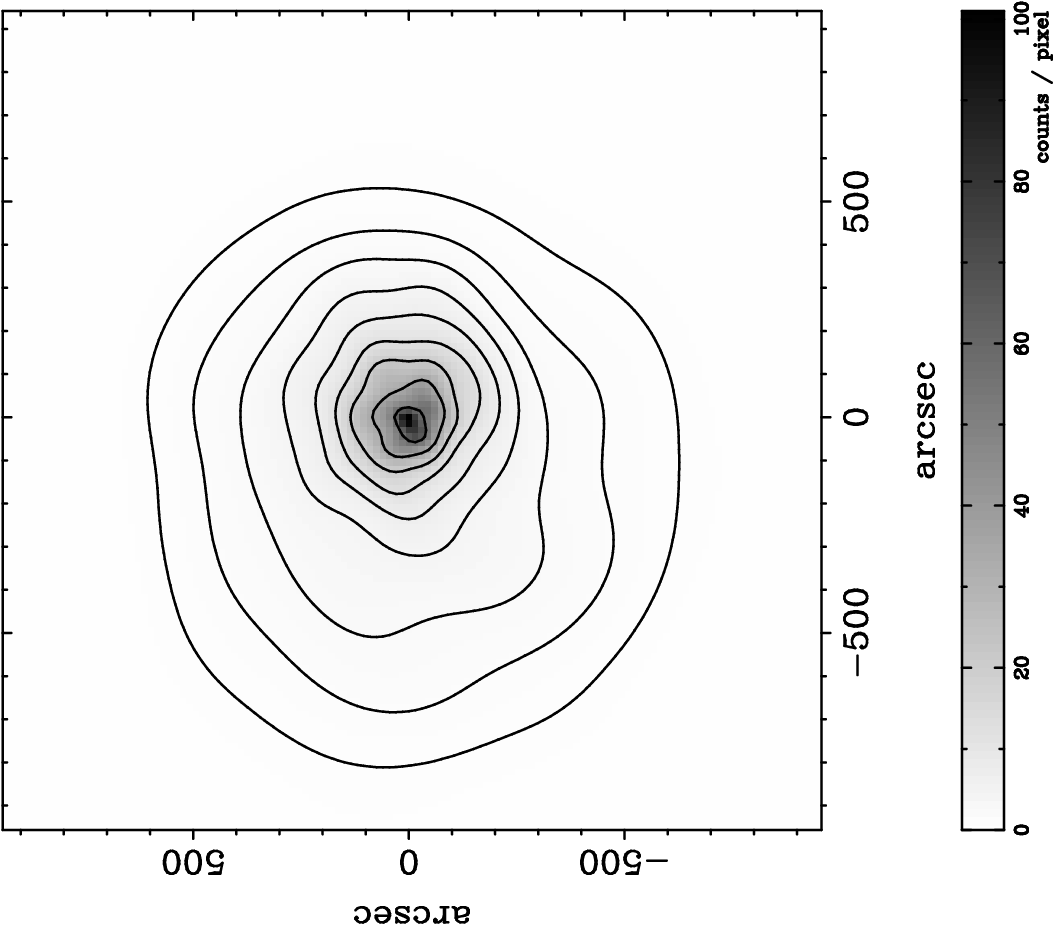}
  \caption{Adaptively-smoothed images of the Perseus Cluster. The
    images are shown in bands B (left), C (middle) and D (right). The
    greyscale is linear and the contours are logarithmic in smoothed
    counts.}
  \label{fig:per_adapt}
\end{figure*}

\begin{figure*}
  \includegraphics[angle=-90,width=0.28\textwidth]{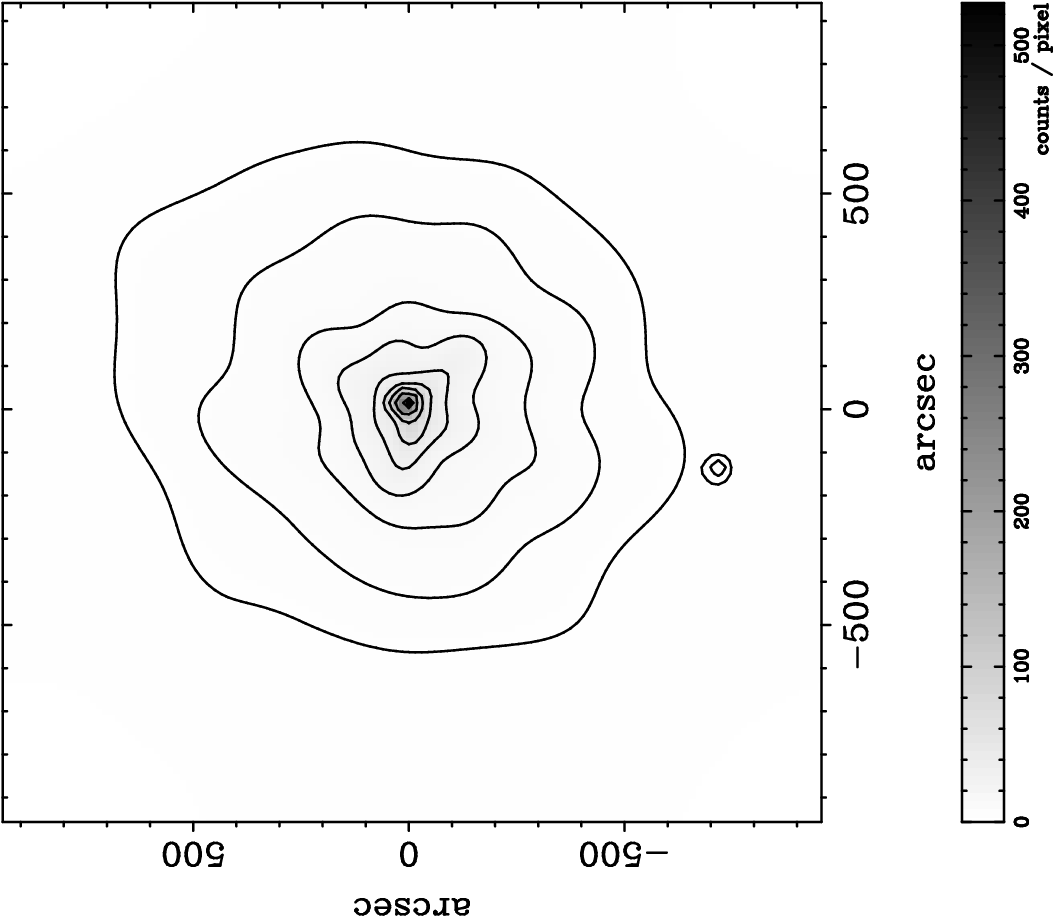}
  \hspace{6mm}
  \includegraphics[angle=-90,width=0.28\textwidth]{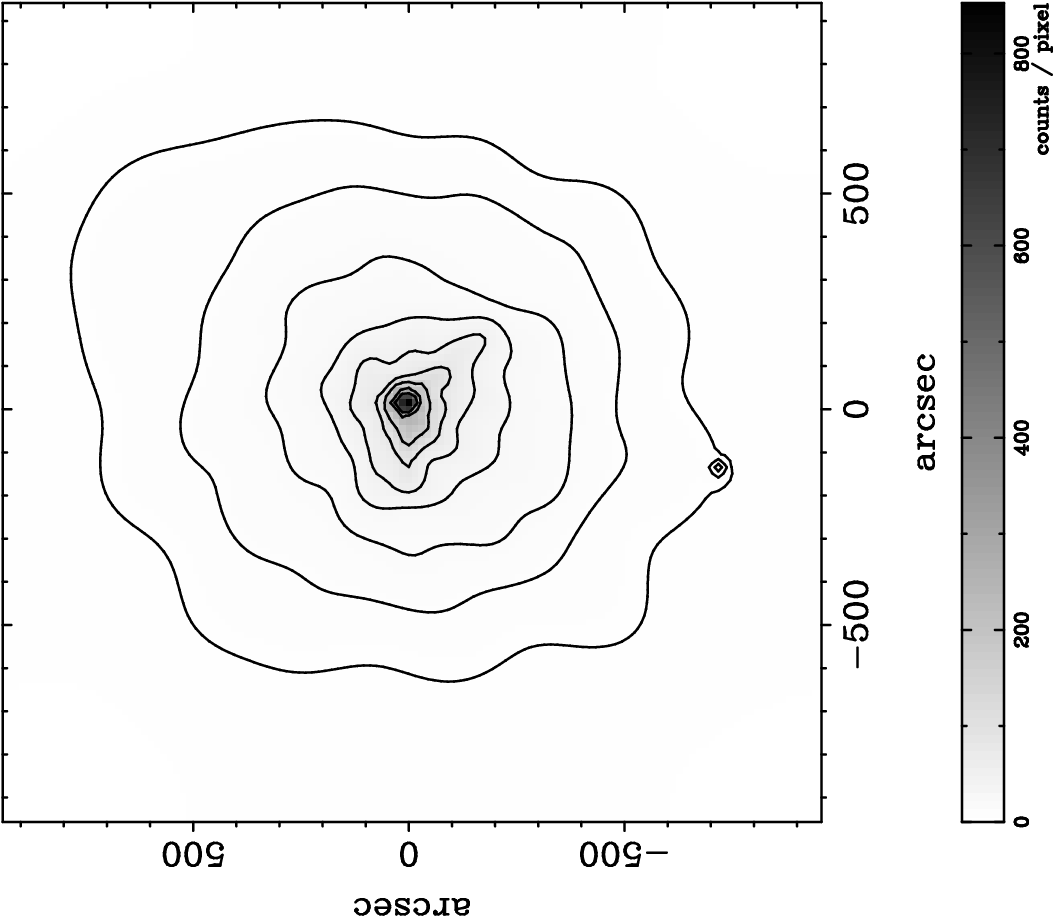}
  \hspace{6mm}
  \includegraphics[angle=-90,width=0.28\textwidth]{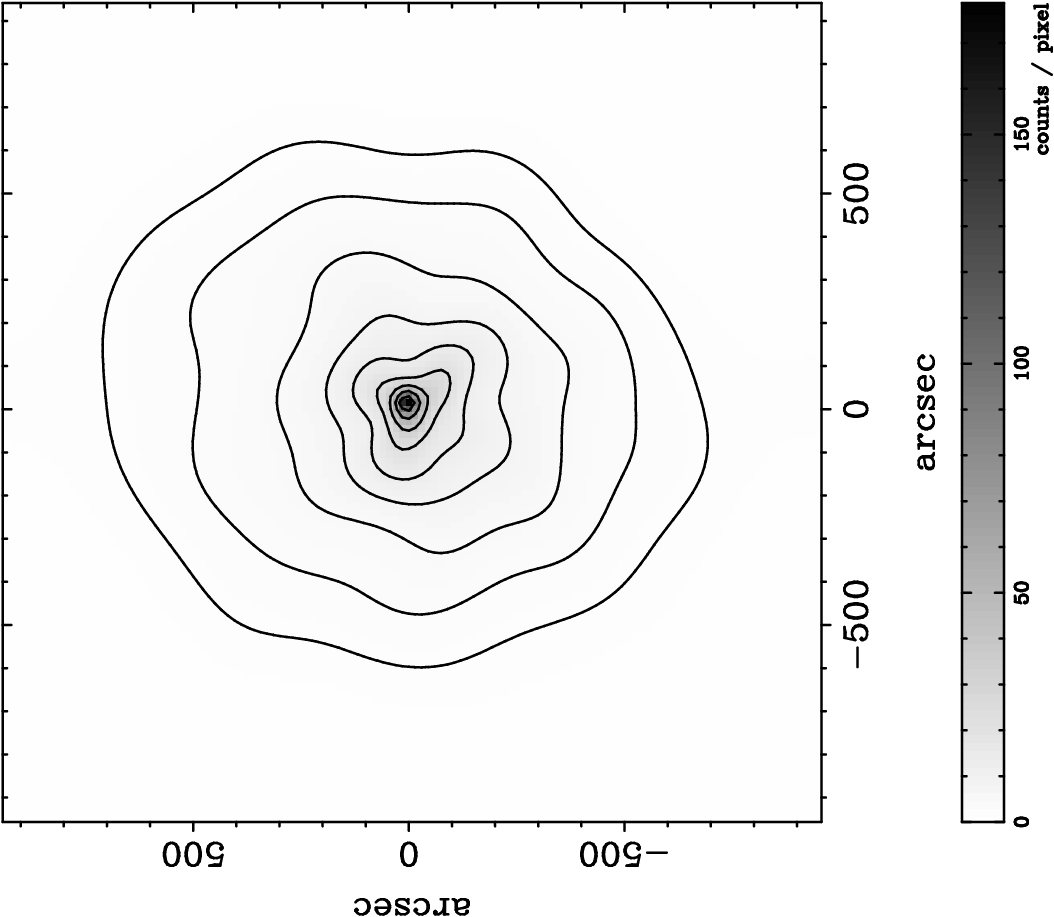}
  \caption{Adaptively-smoothed images of the Virgo Cluster (see
  Fig. \ref{fig:per_adapt} for a description).}
  \label{fig:vir_adapt}
\end{figure*}

\begin{figure*}
  \includegraphics[angle=-90,width=0.28\textwidth]{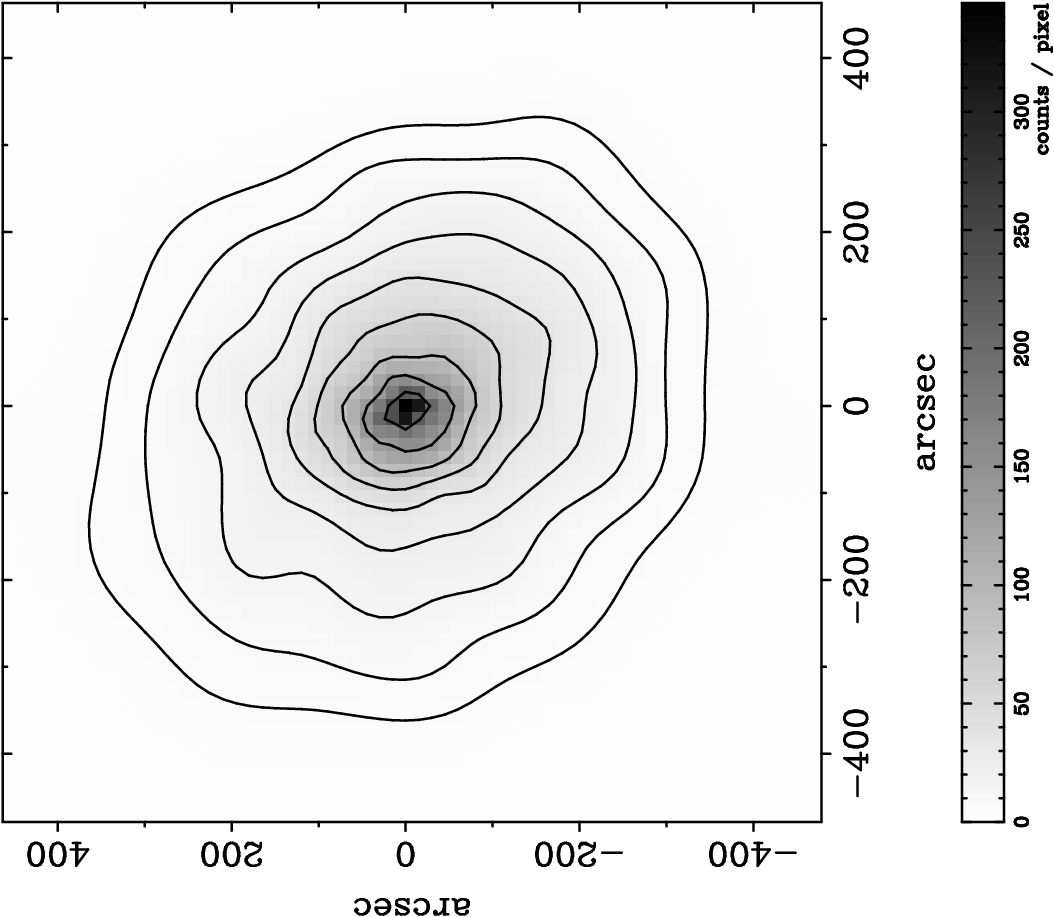}
  \hspace{6mm}
  \includegraphics[angle=-90,width=0.28\textwidth]{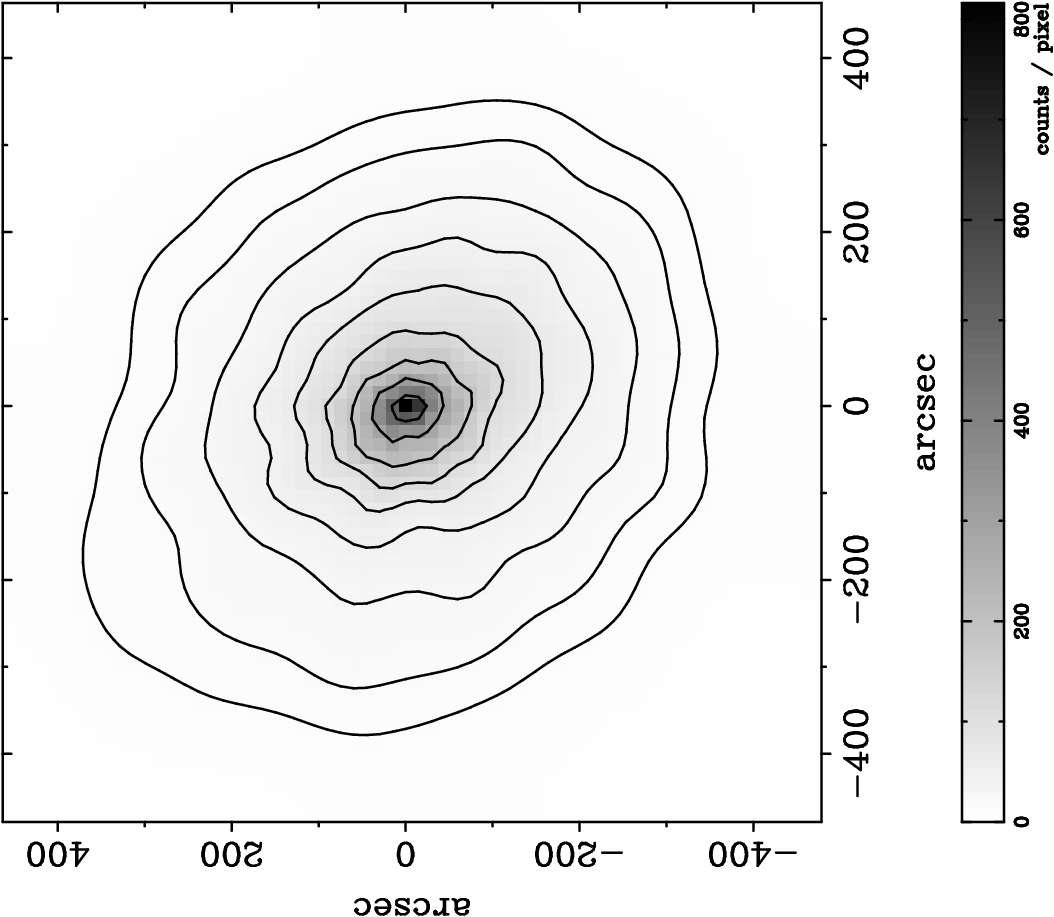}
  \hspace{6mm}
  \includegraphics[angle=-90,width=0.28\textwidth]{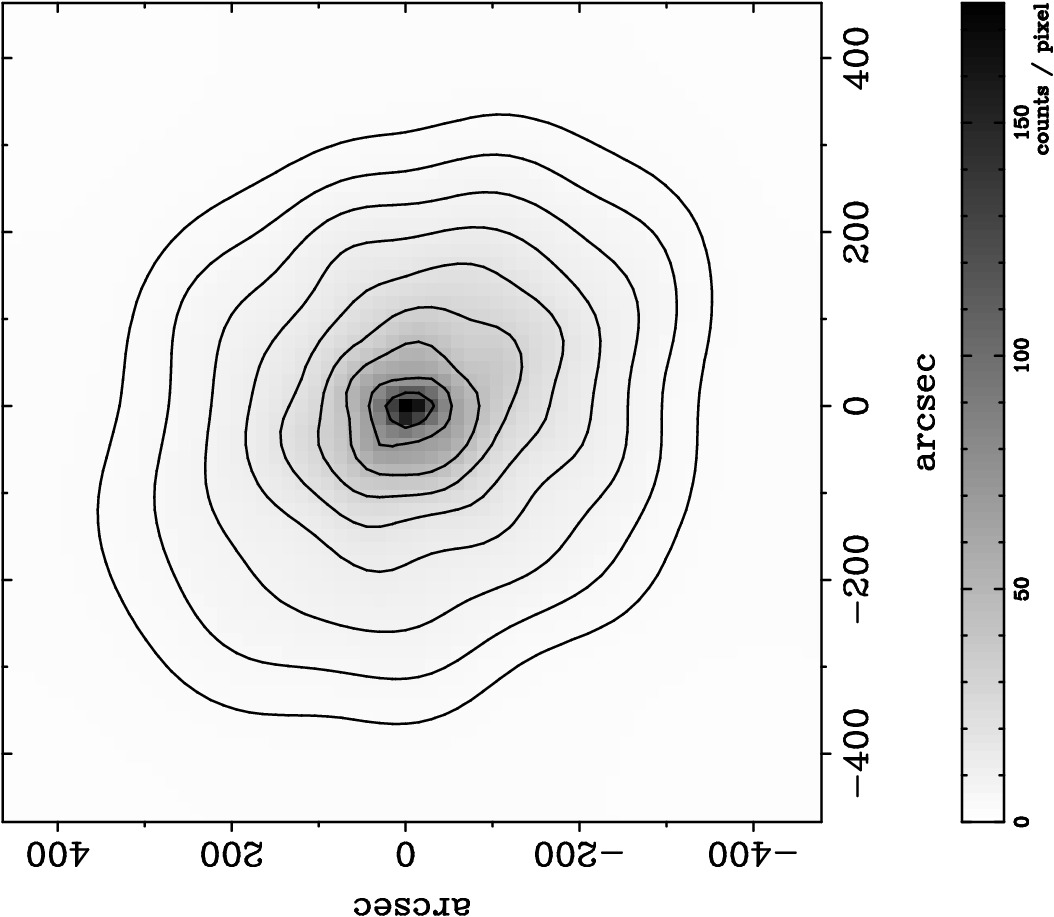}
  \caption{Adaptively-smoothed images of A2199 (see
  Fig. \ref{fig:per_adapt} for a description).}
  \label{fig:a2199_adapt}
\end{figure*}

We created adaptively-smoothed images of the sample of clusters in the
$B$, $C$ and $D$ bands, using an \textsc{asmooth} minimum significance
of 6-$\sigma$. Fig. \ref{fig:per_adapt}, \ref{fig:vir_adapt} and
\ref{fig:a2199_adapt} show the images for the Perseus, Virgo and A2199
clusters, respectively. \emph{ROSAT} HRI images of the centres of the
Perseus cluster, NGC 1275, and the Virgo cluster, M87, have been
examined by B\"ohringer et al. (1993, 1995), respectively. They found
the thermal plasma was displaced by the radio lobes of NGC 1275 at the
centre of the Perseus cluster.

\begin{figure*}
  \includegraphics[angle=-90,width=0.35\textwidth]{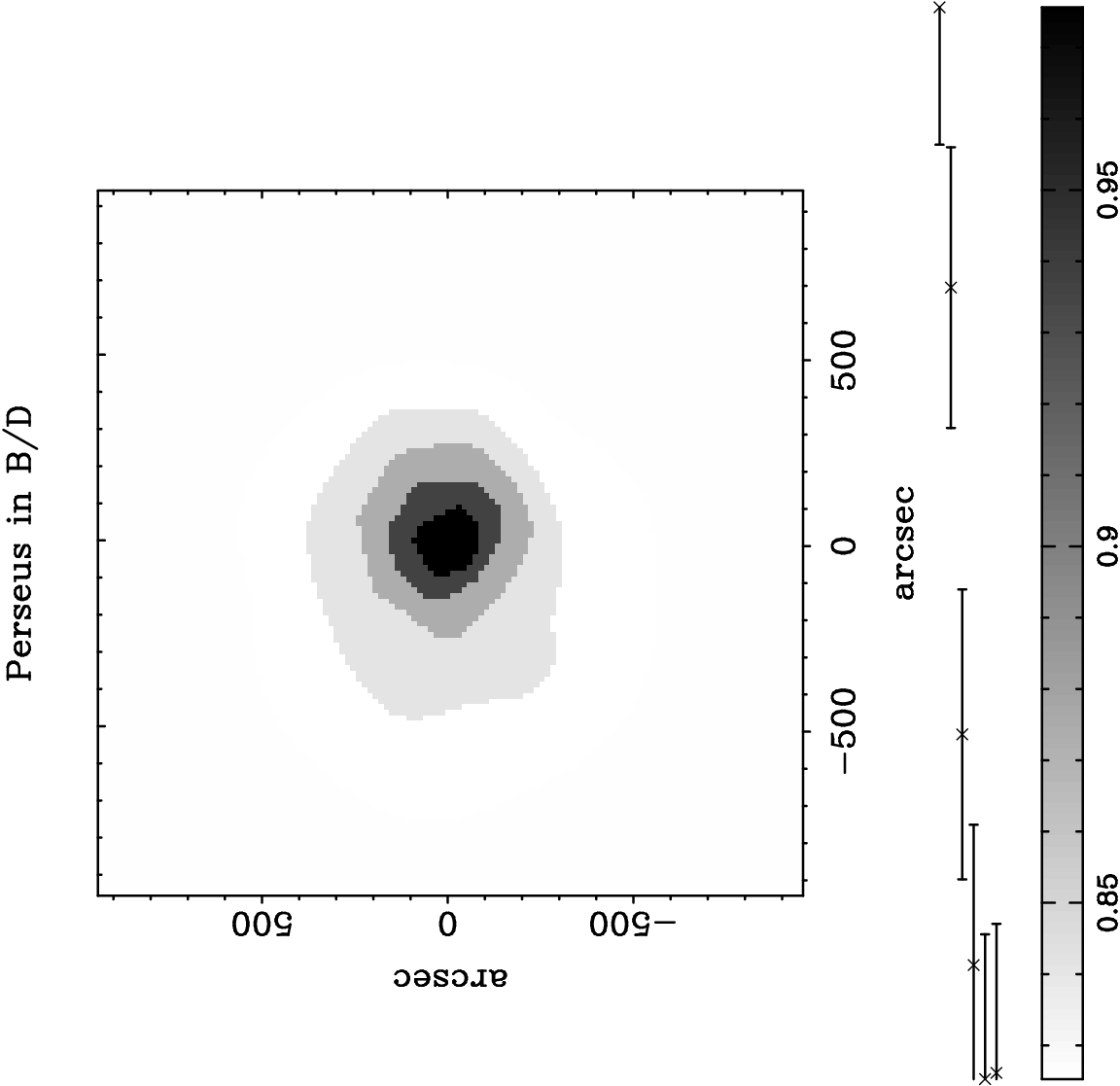}
  \hspace{1cm}
  \includegraphics[angle=-90,width=0.35\textwidth]{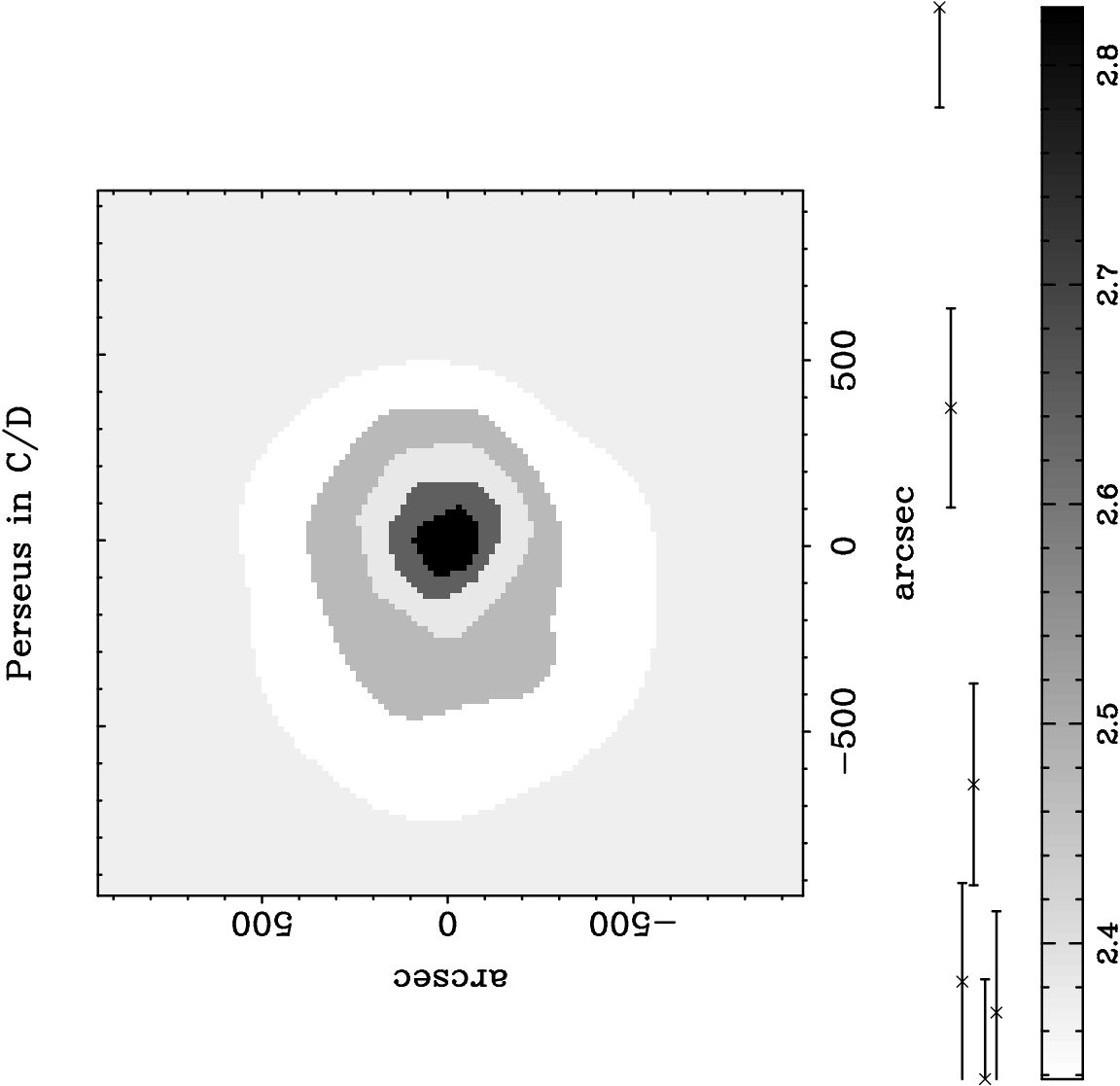}
  \caption{Contour colour maps of the Perseus cluster. Areas between
    contours are coloured by their average X-ray colour. The points
    below the maps show the numerical colours, with their Poisson
    errors. The most central contours are the uppermost points. Darker
    areas indicate softer emission.}
  \label{fig:per_colmap}
\end{figure*}

\begin{figure*}
  \includegraphics[angle=-90,width=0.37\textwidth]{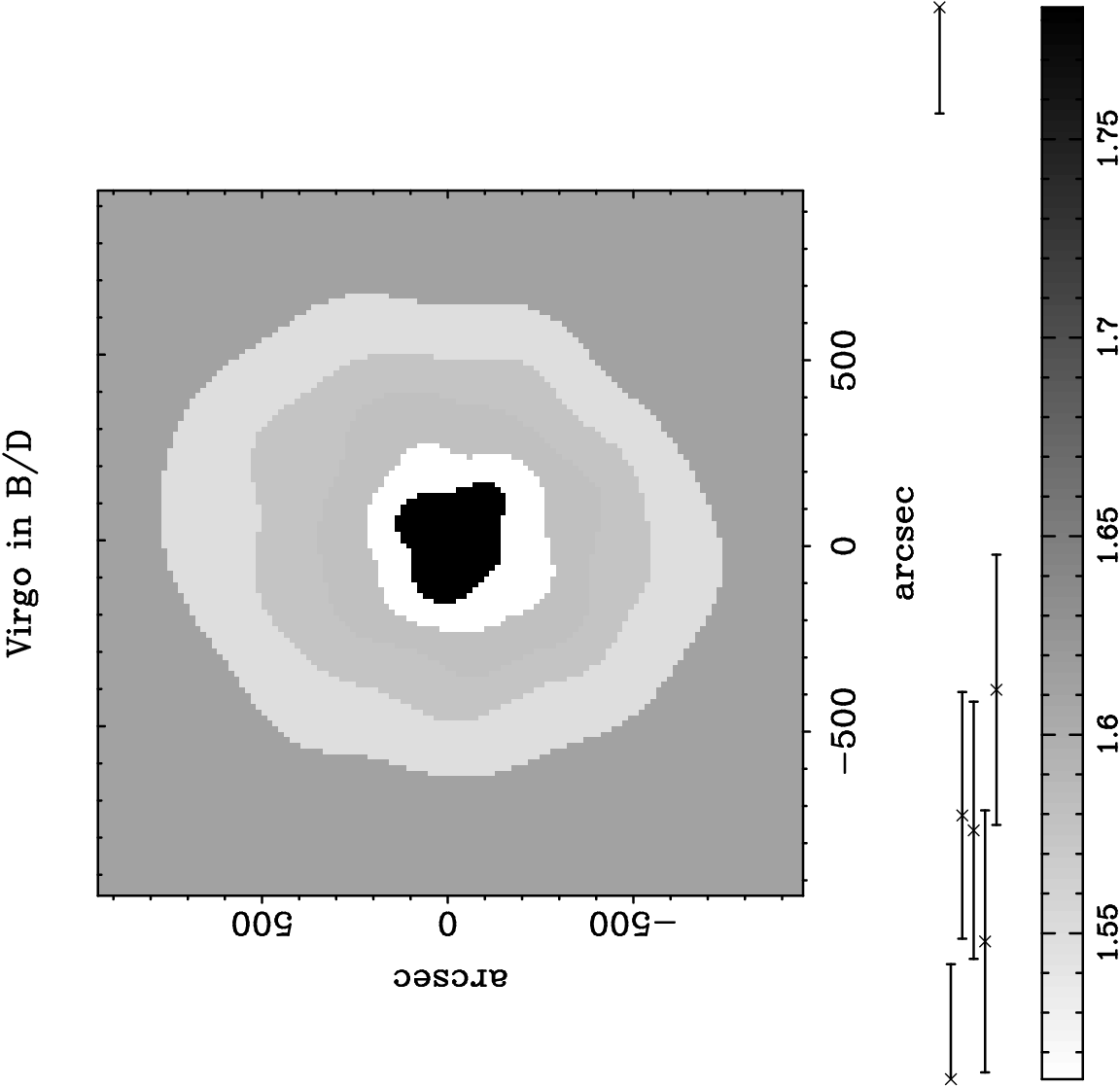}
  \hspace{1cm}
  \includegraphics[angle=-90,width=0.37\textwidth]{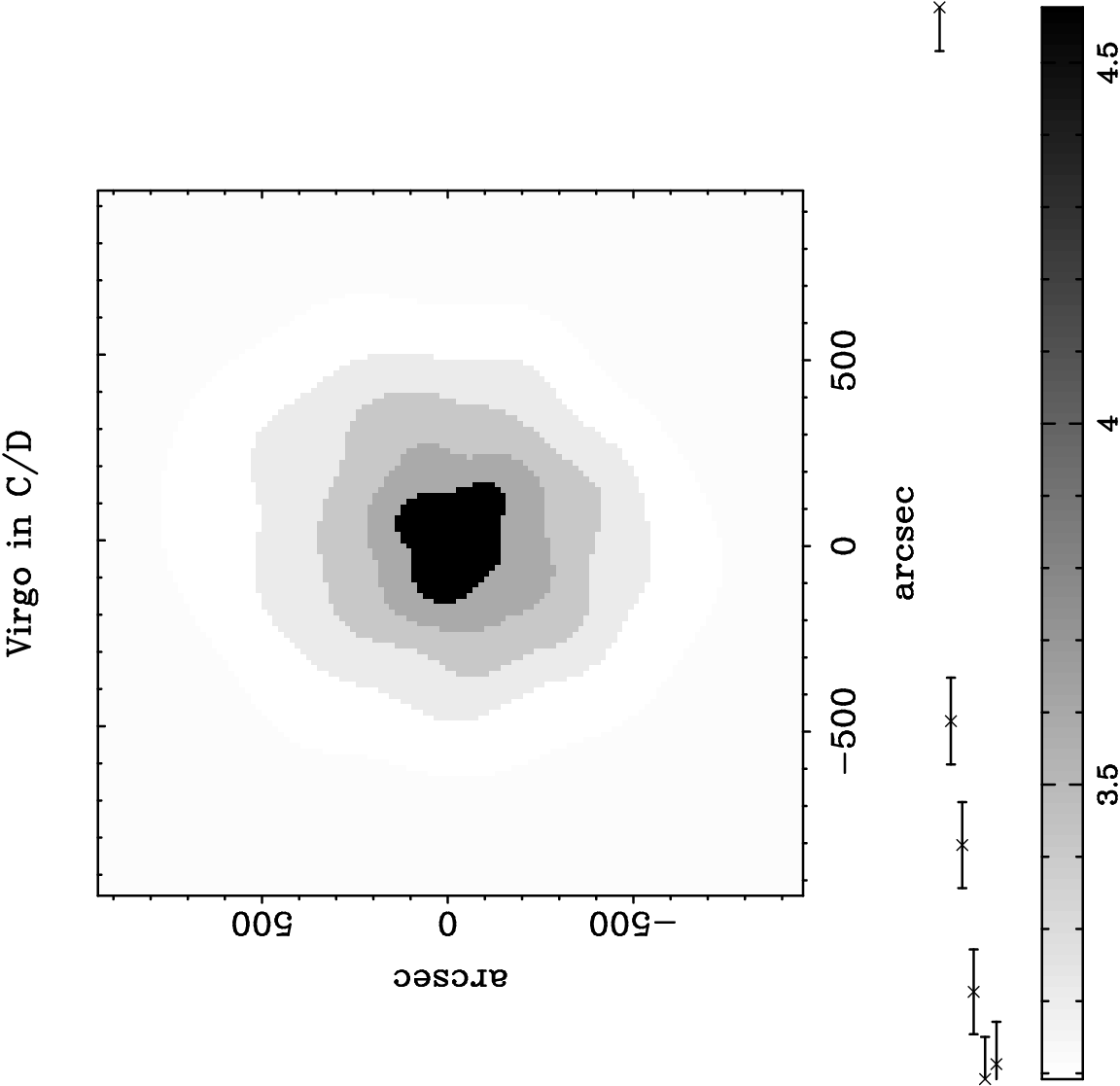}
  \caption{Contour colour maps of the Virgo cluster (see
  Fig. \ref{fig:per_colmap} for a description).}
  \label{fig:vir_colmap}
\end{figure*}

Fig. \ref{fig:per_colmap} and \ref{fig:vir_colmap} are colour contour
maps of the Perseus and Virgo clusters. The maps show the average
X-ray colours between each of six contours. The scale below each map
shows the value and statistical uncertainty of the colour of each
contour. The extreme error bars on the softest and hardest points are
not shown on the scales.  Note that darker shades in these plots
indicate softer emission. Table \ref{tab:disttab} shows the mean physical
distance from the cluster centre of each contour in each cluster in
pixels. It also identifies which clusters were examined using a
$64\times64$ pixel image or a $128\times128$ image.

\begin{figure*}
  \begin{tabular}{ll}
    \includegraphics[angle=-90,width=0.37\textwidth]{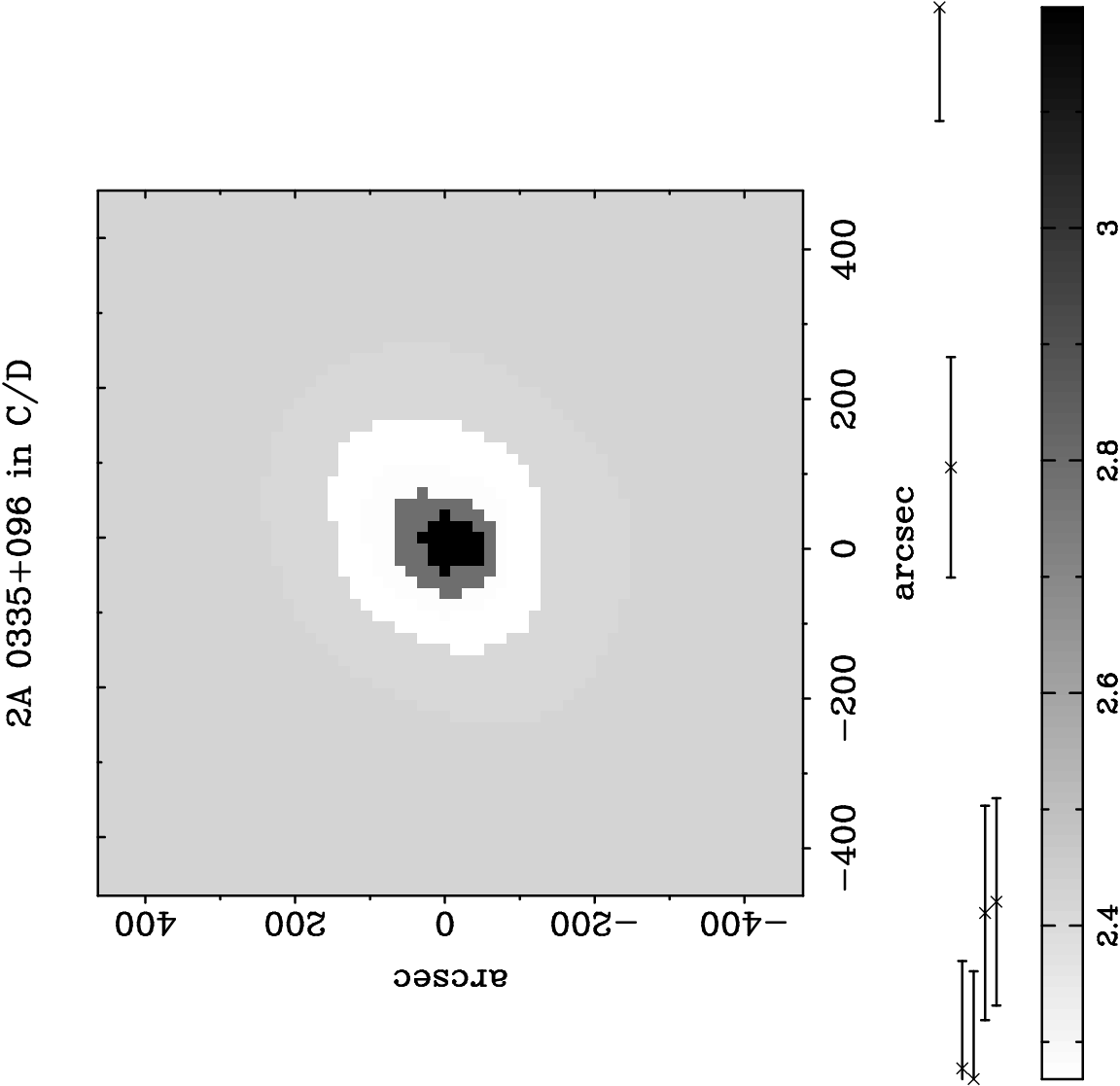} &
    \hspace{5mm}
    \includegraphics[angle=-90,width=0.37\textwidth]{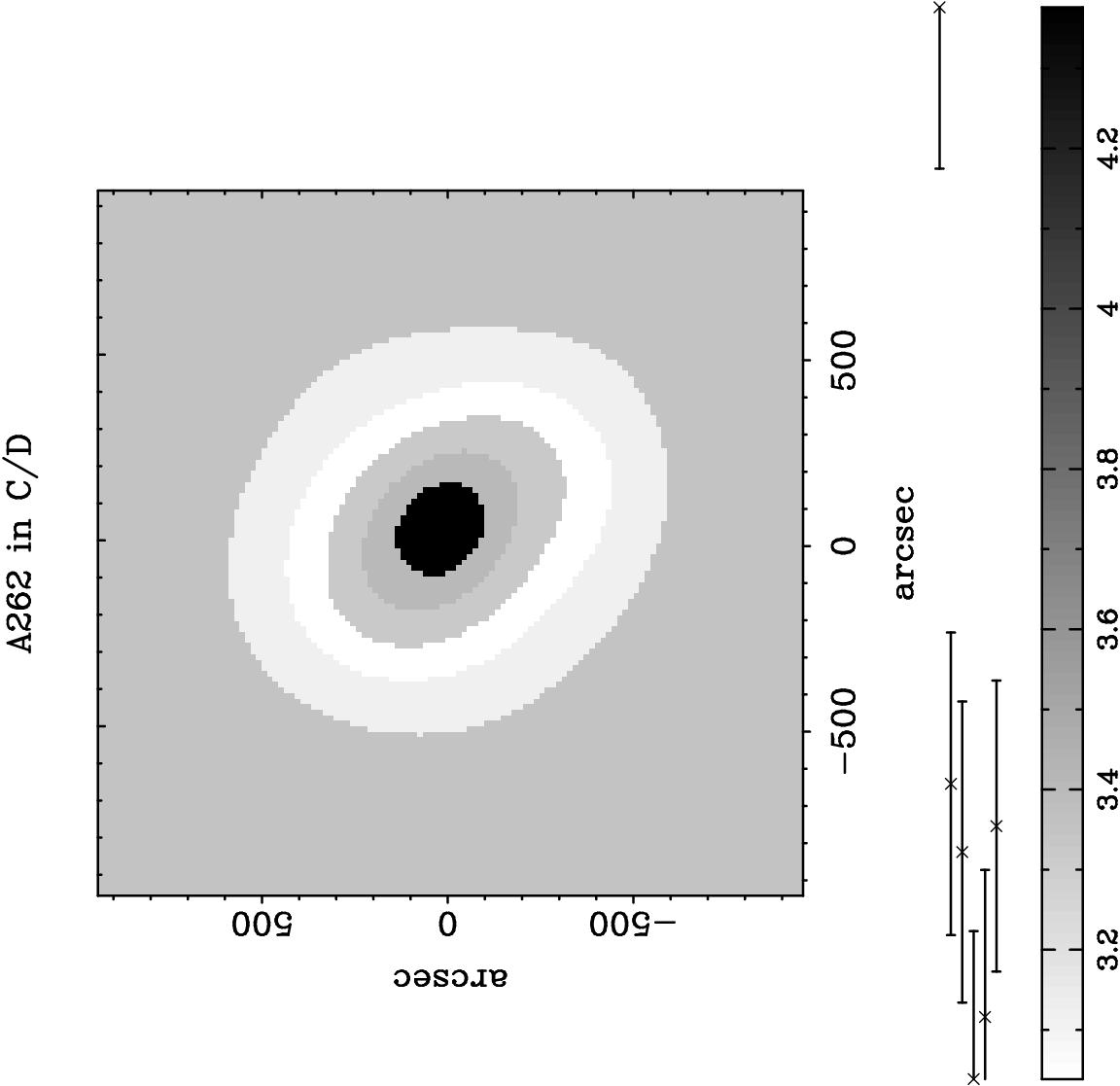}
    \vspace{5mm} \\
    \includegraphics[angle=-90,width=0.37\textwidth]{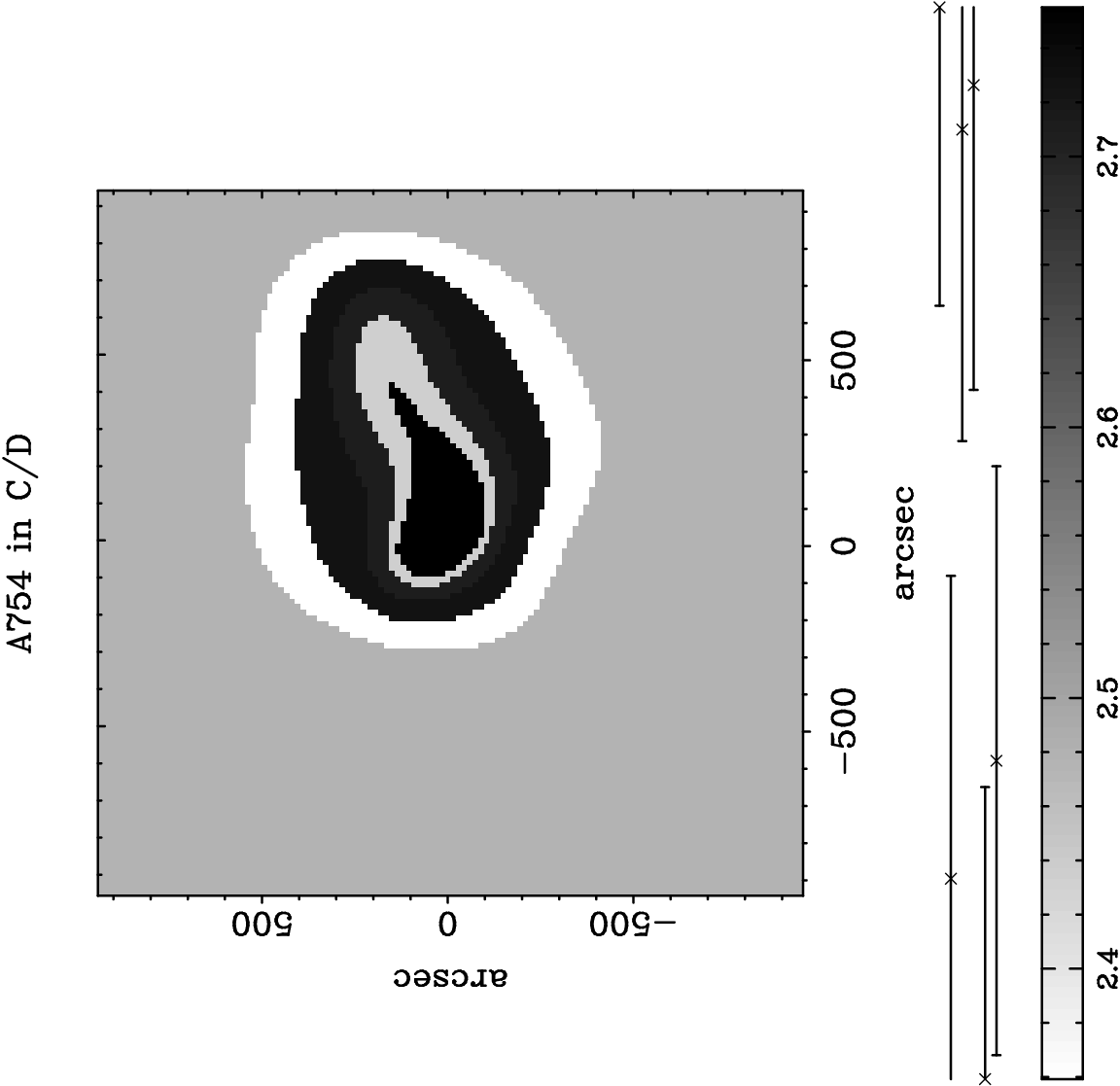} &
    \hspace{5mm}
    \includegraphics[angle=-90,width=0.37\textwidth]{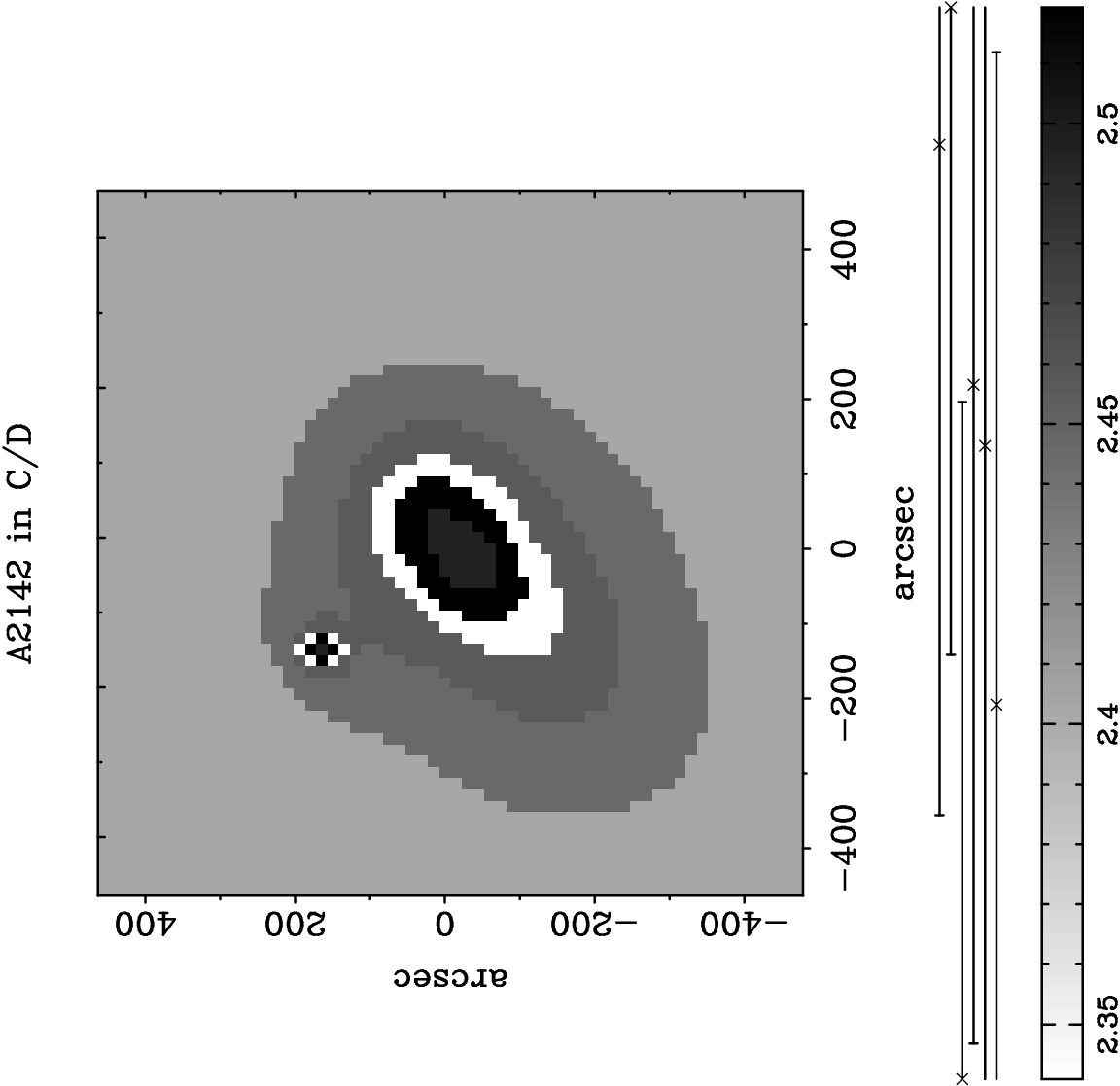}
    \vspace{5mm} \\
    \includegraphics[angle=-90,width=0.37\textwidth]{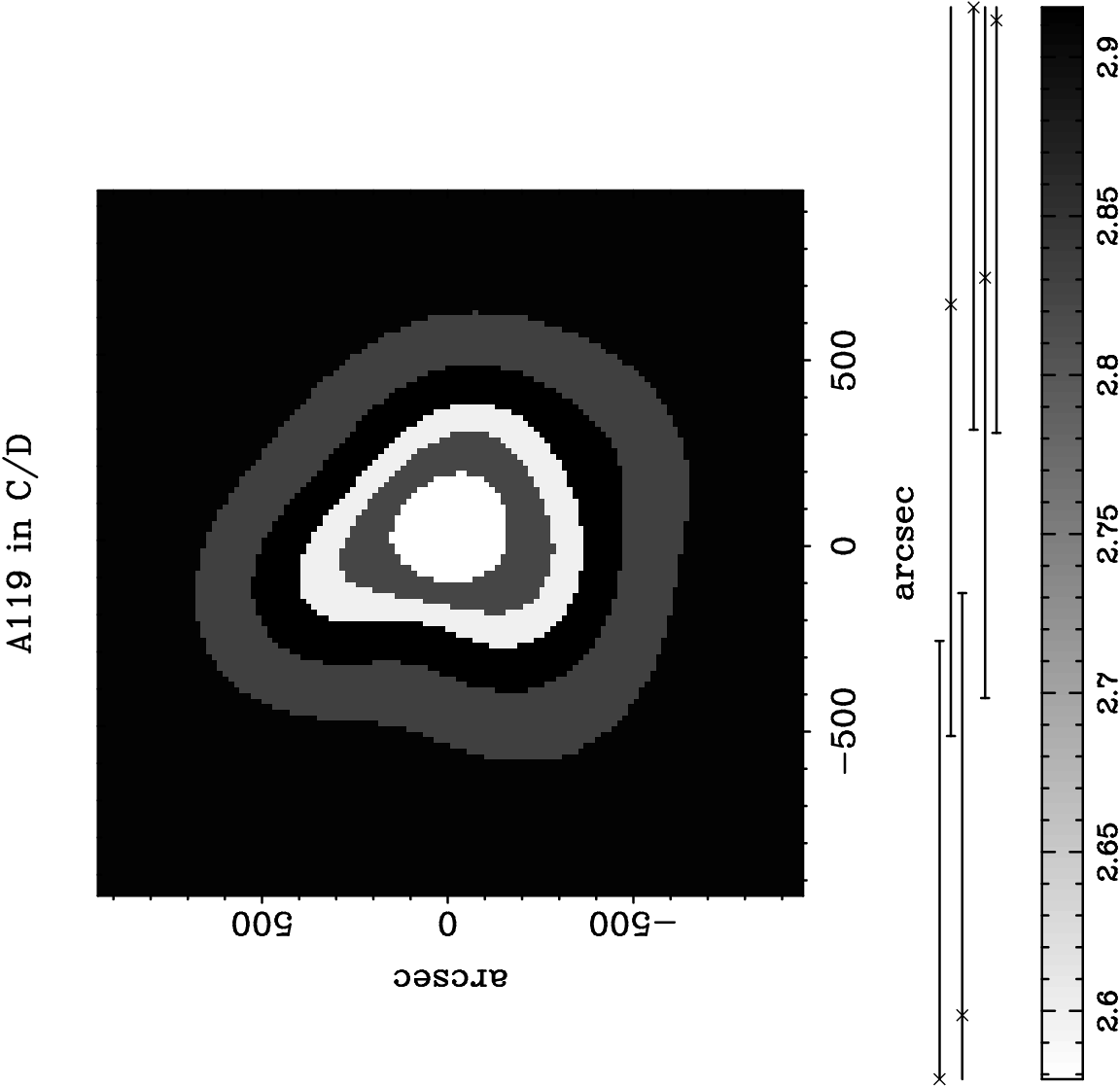} &
    \hspace{5mm}
    \includegraphics[angle=-90,width=0.37\textwidth]{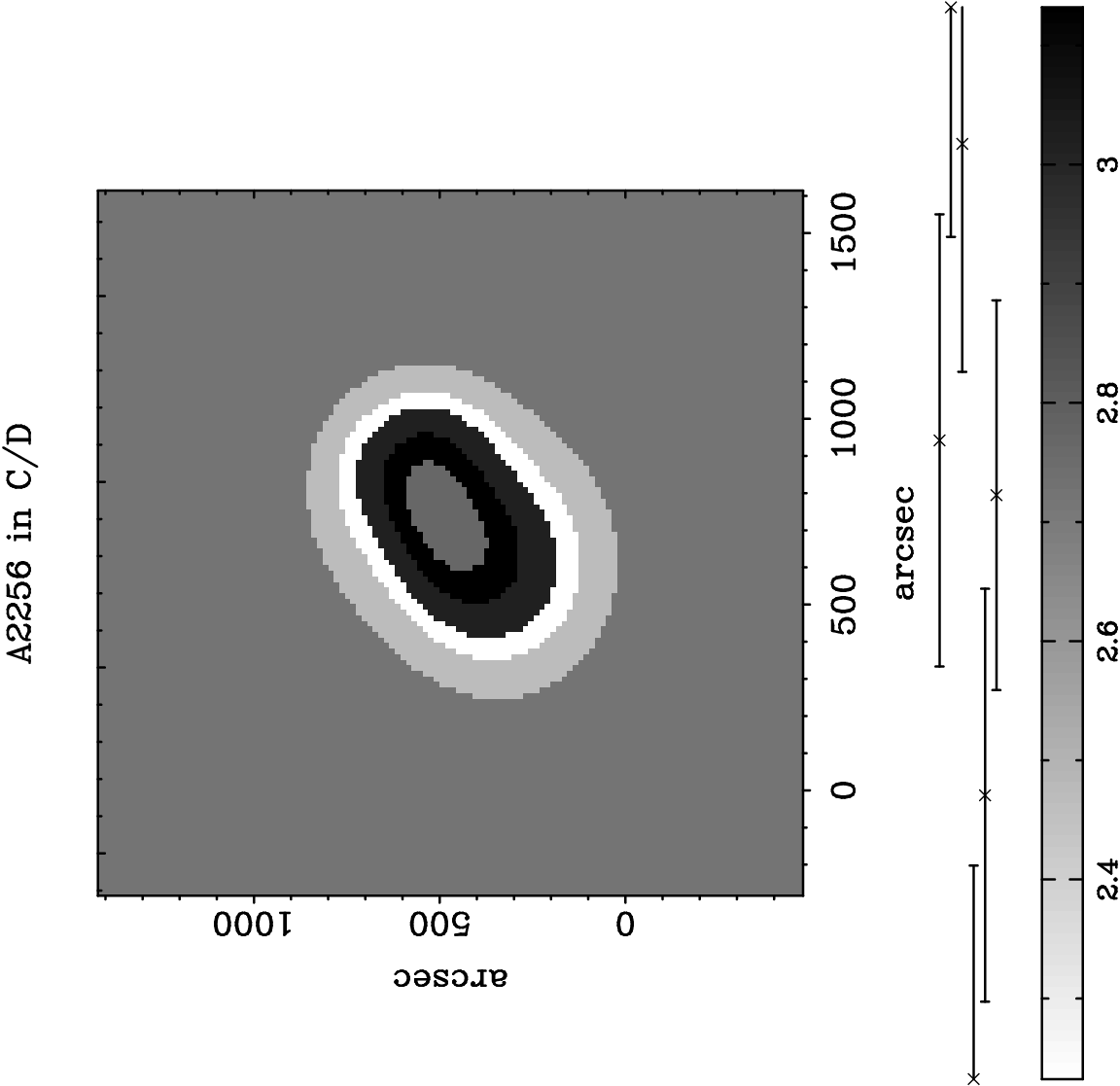} \\
  \end{tabular}
  \caption{A selection of $C/D$ contour colour maps (see
    Fig. \ref{fig:per_colmap} for a description). The upper two
    clusters contain cooling flows.  The others are non-cooling-flow,
    or low $\dot{M}$ cooling flow, clusters.}
  \label{fig:cdset_colmap}
\end{figure*}

Fig. \ref{fig:cdset_colmap} shows a sample of six colour contour
maps. Four of the clusters are non-cooling-flow clusters, showing
interesting features. Two are cooling flow clusters, which make up
most of our sample, and they reveal little substructure. The point
source in A2142 is shown, but is not included in the contour
statistics.

\begin{figure*}
  \includegraphics[width=0.88\textwidth]{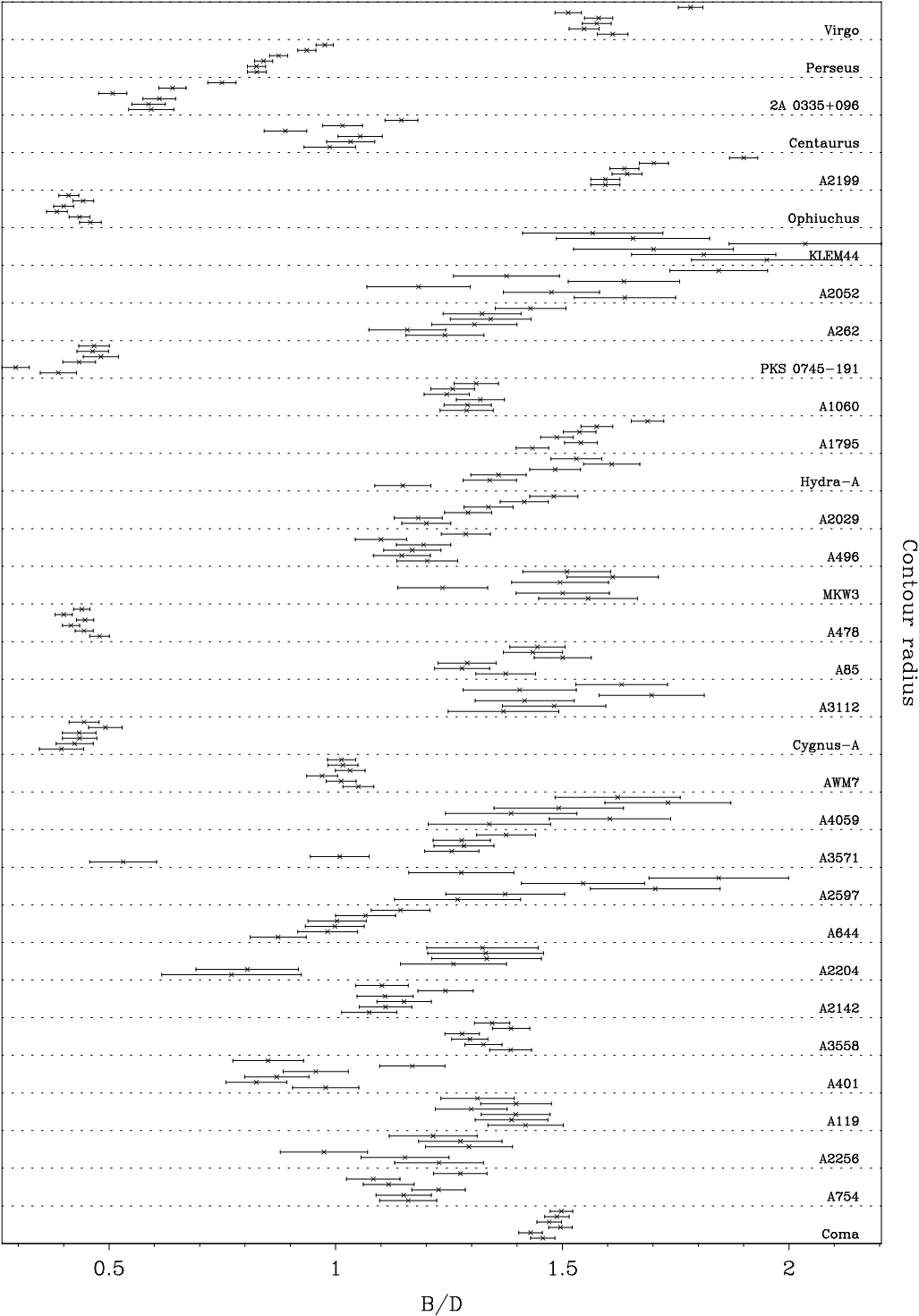}
  \caption{Variation of the ratio B/D for the sample of clusters. Each
    point represents the colour of a contour. The uppermost points
    represent the innermost contours, numbered 1.}
  \label{fig:b_d_corr}
\end{figure*}

\begin{figure*}
  \includegraphics[width=0.88\textwidth]{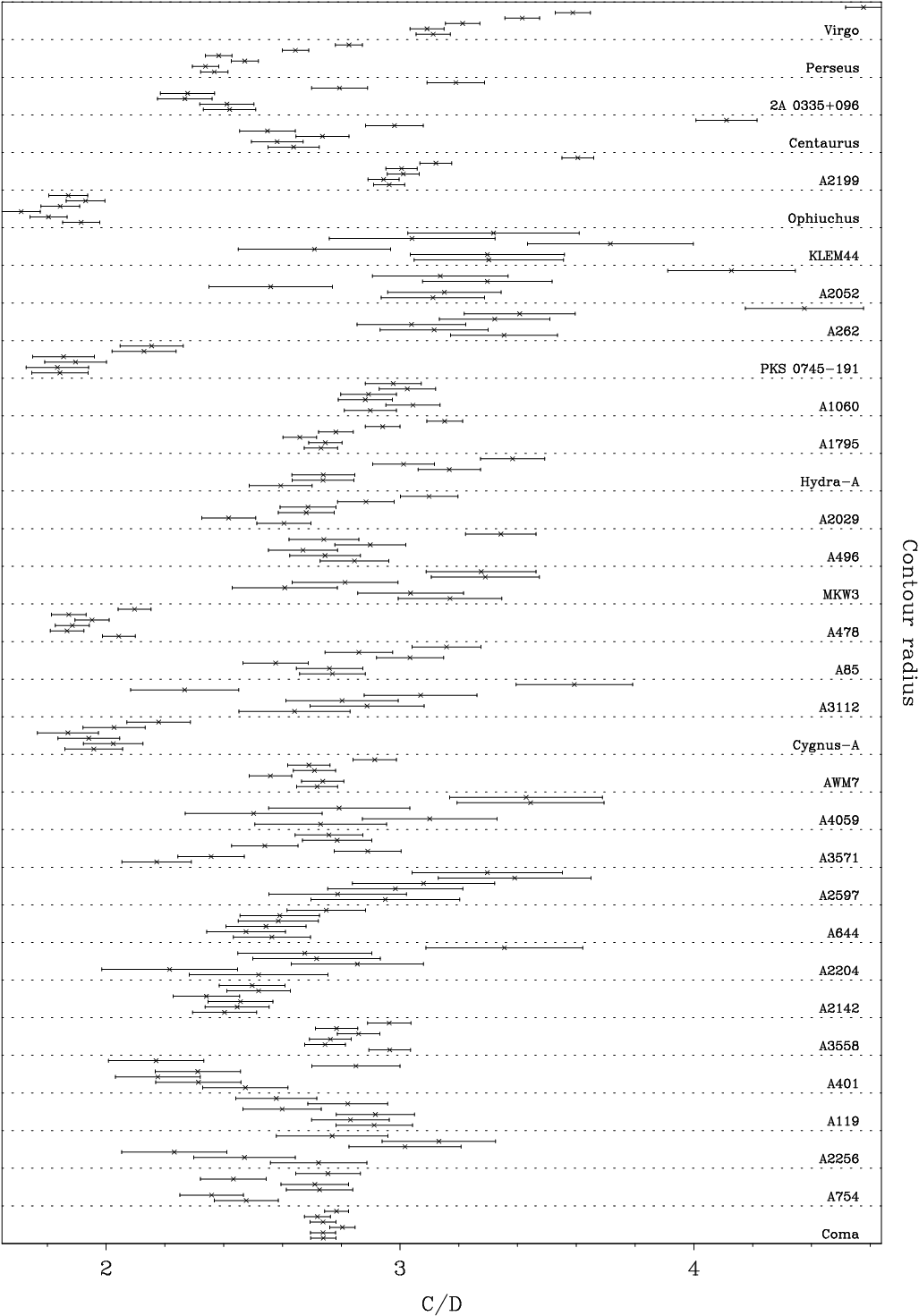}
  \caption{Variation of the ratio C/D for the sample of clusters. Each
    point represents the colour of a contour. The uppermost points
    represent the innermost contours, numbered 1.}
  \label{fig:c_d_corr}
\end{figure*}

Fig. \ref{fig:b_d_corr} and \ref{fig:c_d_corr} show the average $B/D$
and $C/D$ colours for the contours of each of the sample of
clusters. The clusters are listed in decreasing cooling flow flux
order ($\dot{M}/z^2$, where $\dot{M}$ is the mass deposition rate, and
$z$ is the redshift of the cluster) using the PSPC $\dot{M}$ values
obtained by Peres et al. (1998). However, some of the clusters have
highly uncertain mass deposition rates, so the ordering is not
definite.

Table \ref{tab:clusters} shows a significance measure of the change
in colour from the outside to the inner region for each cluster. For a
cluster with contours of colour $X_i$, it is calculated using
\begin{equation}
  \label{eqn:sig}
  S(X) = \frac{1}{\sqrt{ 3/2 } \: \sigma_X} \left[ X_1 - \left(
    X_5+X_6 \right)/2 \right],
\end{equation}
$X_1$ is the innermost colour value, $X_5$ and $X_6$ are the two
outermost colours, $\sigma_X$ is the statistical uncertainty in each
colour. The uncertainties are almost the same for each contour, due to
our method of choosing contour levels (Section \ref{sec:obs}).
$\sqrt{3/2}$ is a numerical factor to
make $S(X)$ a measure of the number of 1-$\sigma$ errors. We take the
mean outer two colours to decrease the chance of background
contamination. The background is a large fraction of the total number
of counts in the outer contour for several clusters, for example
A3571, A2204, A3558 and A262.

\begin{figure}
  \includegraphics[angle=-90,width=0.90\columnwidth]{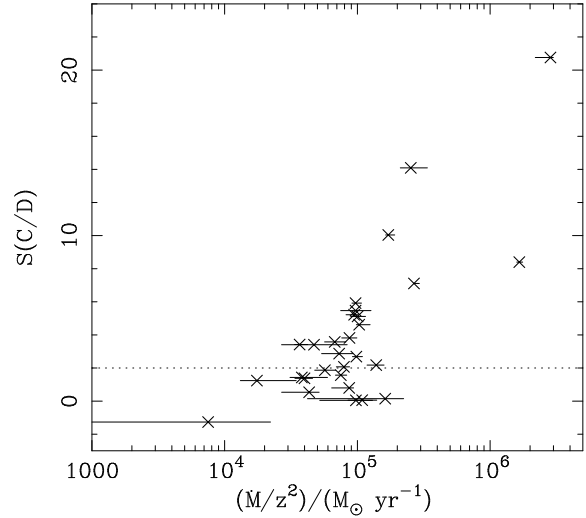}
  \caption{Plot of cooling flow flux, $\dot{M}/z^2$, against the
    significance measure of a colour gradient $S(C/D)$ (Section
    \ref{sec:results}). The 2-$\sigma$ significance level is indicated
    by a dotted line.}
  \label{fig:colourvsmdotonz}
\end{figure}

\begin{figure}
  \includegraphics[angle=-90,width=0.90\columnwidth]{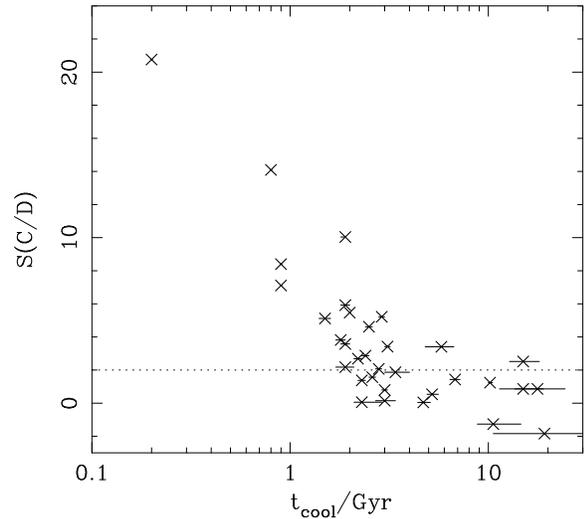}
  \caption{Plot of the central cooling time, $t_{\mathrm{cool}}$,
    against the significance measure of a colour gradient $S(C/D)$
    (Section \ref{sec:results}). The 2-$\sigma$ significance level is
    indicated by a dotted line.}
  \label{fig:colourvstcool}
\end{figure}

Fig. \ref{fig:colourvsmdotonz} shows the variation of the significance
of the $C/D$ colour, presumably due to the presence of cool gas, with
the cooling flow flux, $\dot{M}/z^2$, for the sample of
clusters. Fig. \ref{fig:colourvstcool} shows the significance plotted
against central cooling time, $t_{\mathrm{cool}}$. There is a clear
anti-correlation between $t_{\mathrm{cool}}$ and the significance of a
$C/D$ colour gradient. Clusters with $t_{\mathrm{cool}} \le 2$ Gyr
show colour gradients with $S(C/D)>2$.  $t_{\mathrm{cool}}$ is a
measure of the quality of data for a cooling flow. Cooling flow
clusters with short cooling times have enough counts to make their
central bins small in area.

Some of the results obtained are dependent on our choice of
significance measure. If the cluster has a large soft core, as A3571,
Hydra-A and A2204 seem to have in $B/D$, then using a measure based on
$X_3+X_4$ instead of $X_5+X_6$ in equation (\ref{eqn:sig}) gives a
much lower significance value. A significance based on $X_3+X_4$ would
be less affected by background noise, but throws away much of the
data. A more complex significance measure might be useful, but for
those clusters where it might be important, noise is probably a more
limiting factor.

\begin{figure*}
  \begin{tabular}{llll}
  (a) & \includegraphics[angle=-90,width=0.36\textwidth]{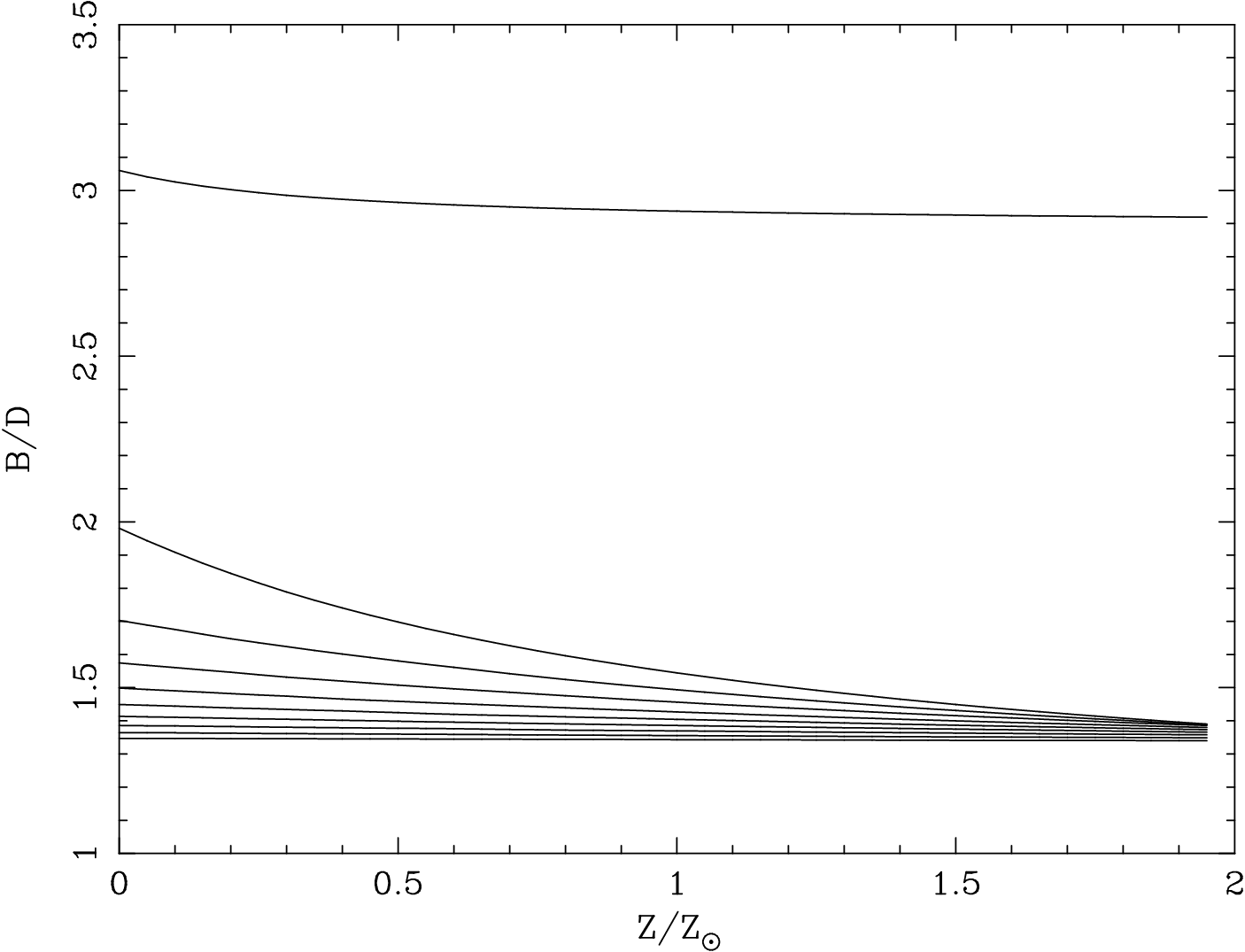}
  & (b) &
  \includegraphics[angle=-90,width=0.36\textwidth]{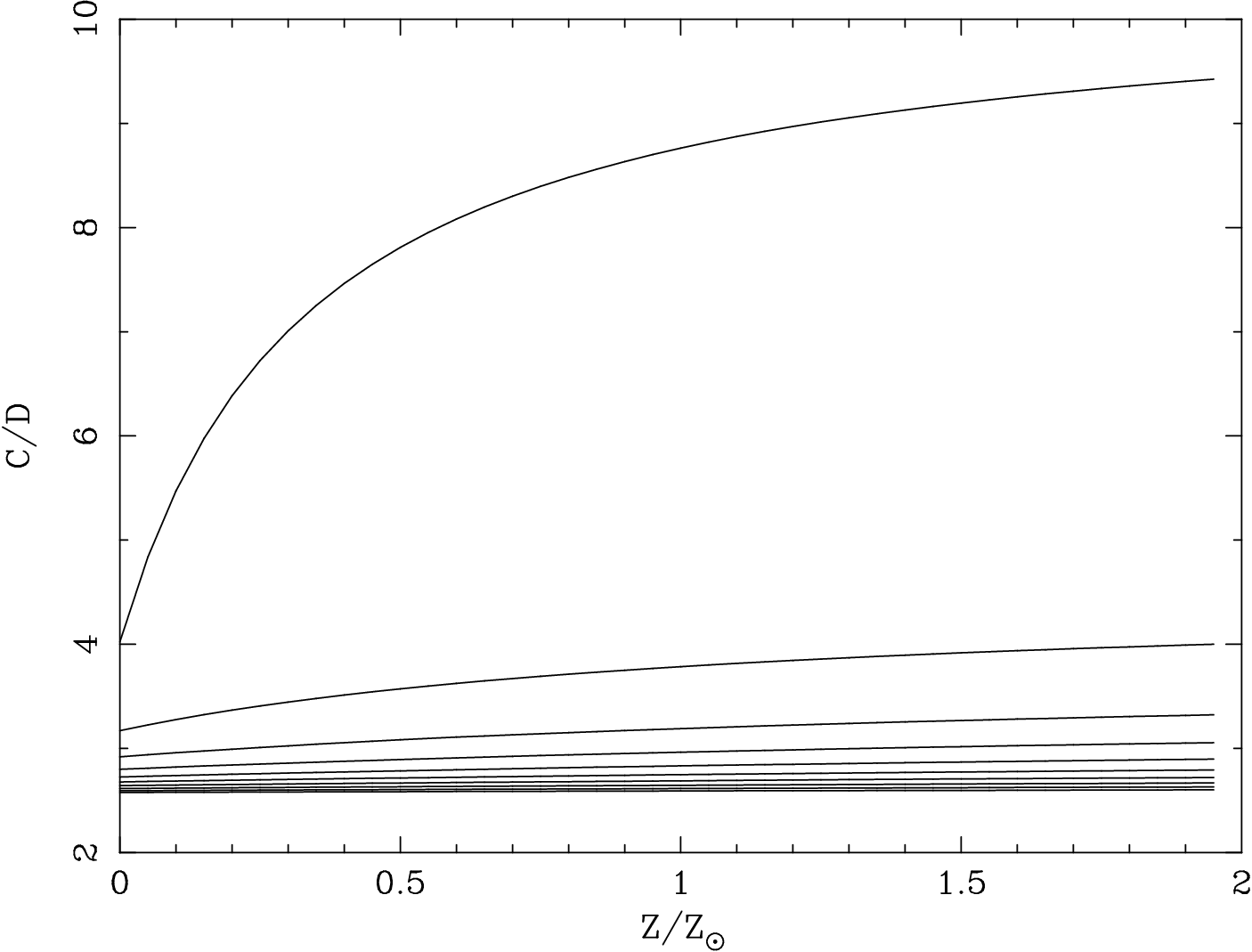} \\
  (c) & \includegraphics[angle=-90,width=0.36\textwidth]{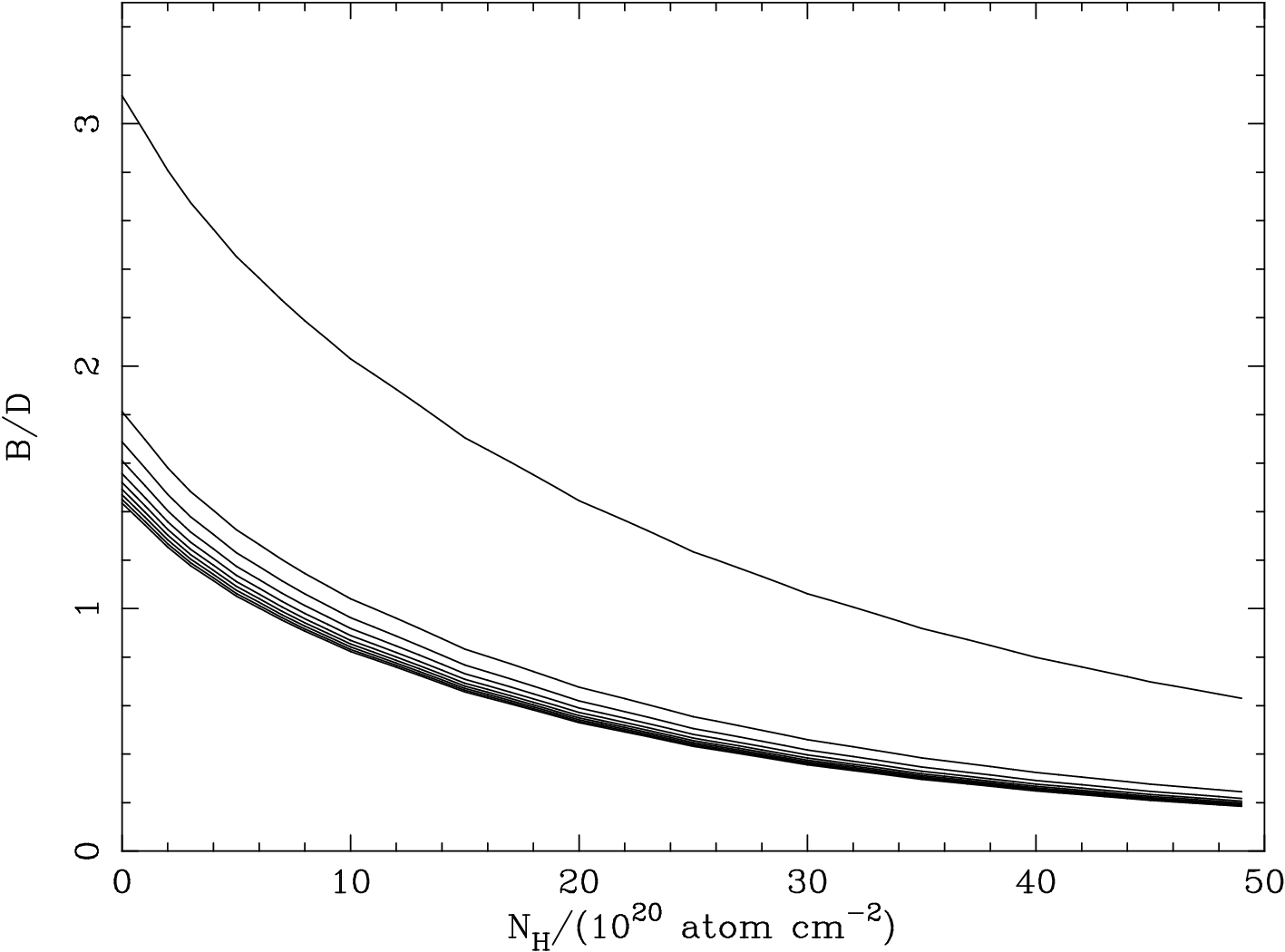}
  & (d) & \includegraphics[angle=-90,width=0.36\textwidth]{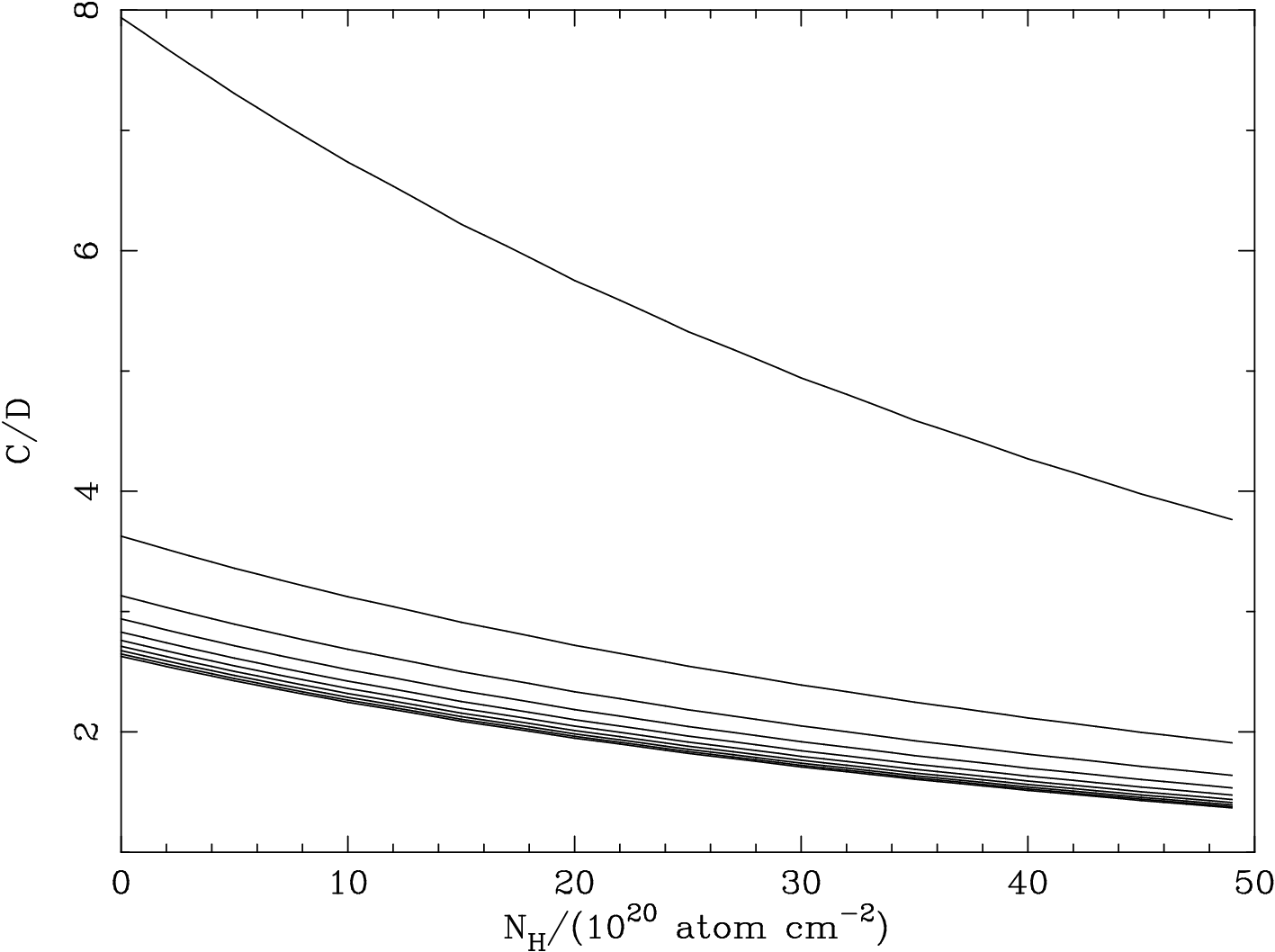}
  \end{tabular}
  \caption{Theoretical predictions for the $B/D$ and $C/D$ ratios for
  an isothermal gas at temperatures of 1--10 keV. The lines show steps
  of 1 keV in temperature. The lowest temperature at the highest ratio
  on the $y$-axis. (a) $B/D$ and (b) $C/D$ as functions of metallicity
  at fixed column density ($N_H = 10^{20} \textrm{ atom
  cm}^{-2}$). (c) $B/D$ and (d) $C/D$ as functions of column density
  at fixed metallicity ($Z = 0.5 Z_{\odot}$).}
\label{fig:isomodels}
\end{figure*}

\begin{figure*}
  \begin{tabular}{llll}
  (a) & \includegraphics[angle=-90,width=0.36\textwidth]{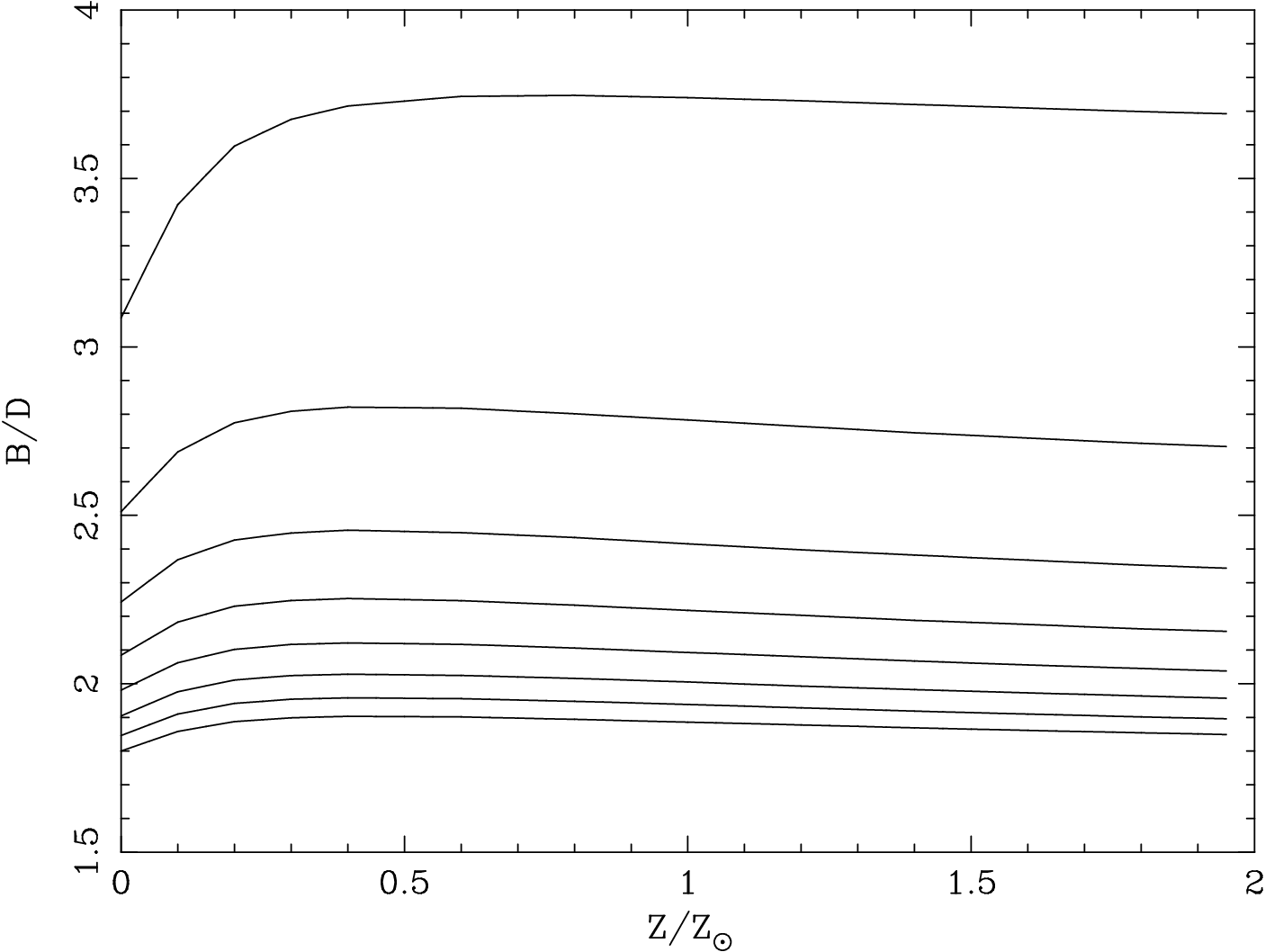}
  & (b) &
  \includegraphics[angle=-90,width=0.36\textwidth]{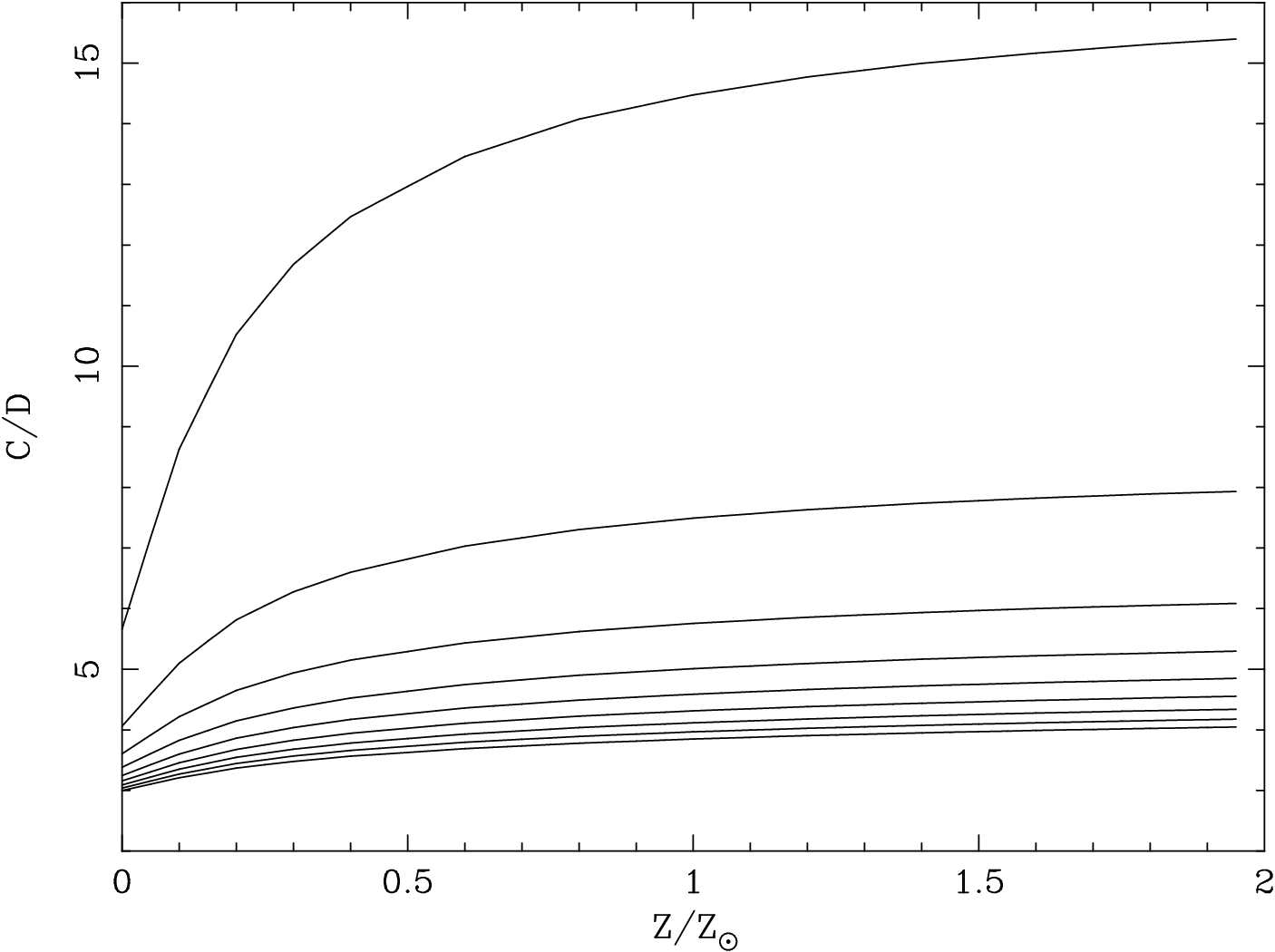} \\
  (c) & \includegraphics[angle=-90,width=0.36\textwidth]{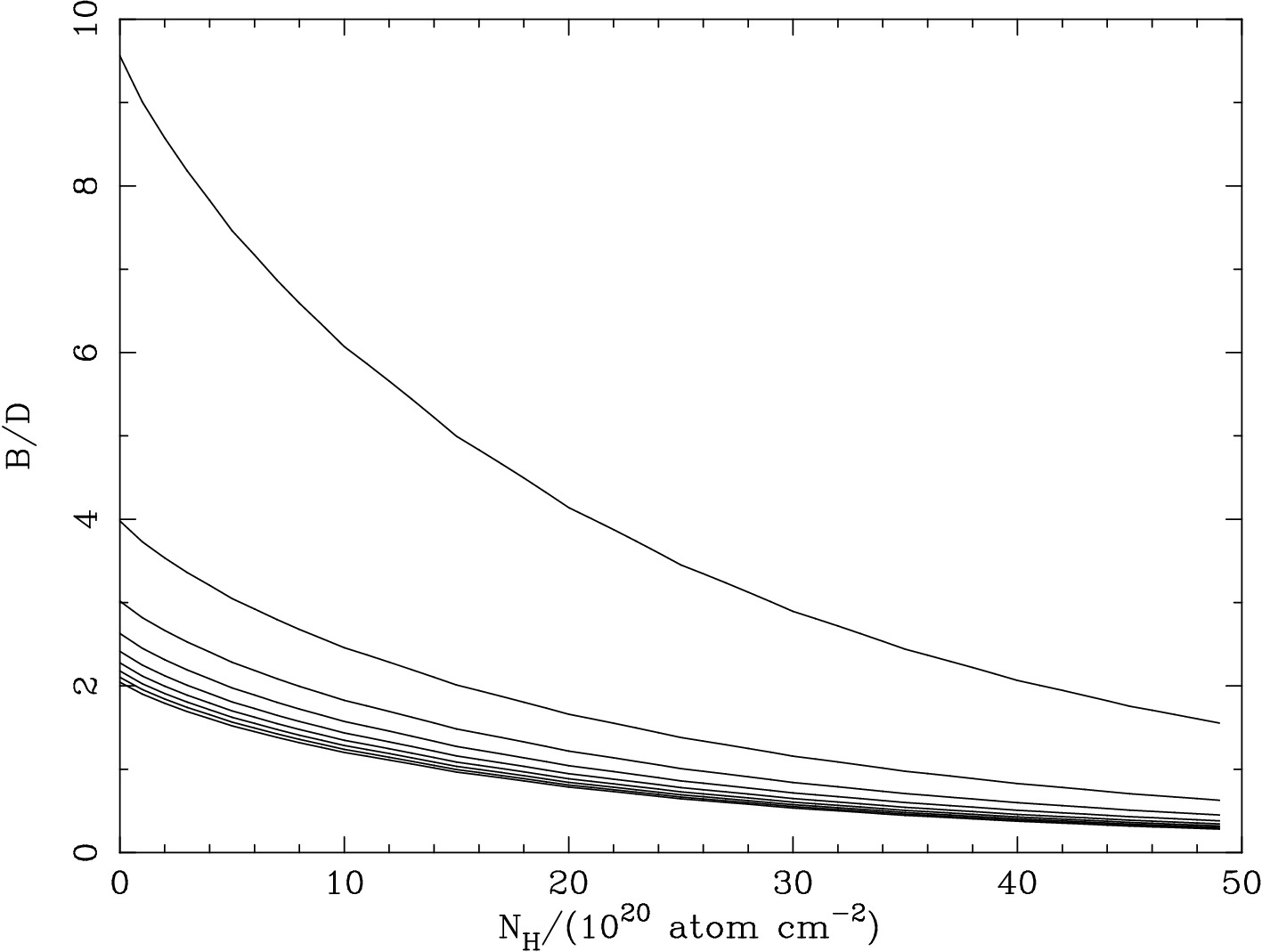}
  & (d) & \includegraphics[angle=-90,width=0.36\textwidth]{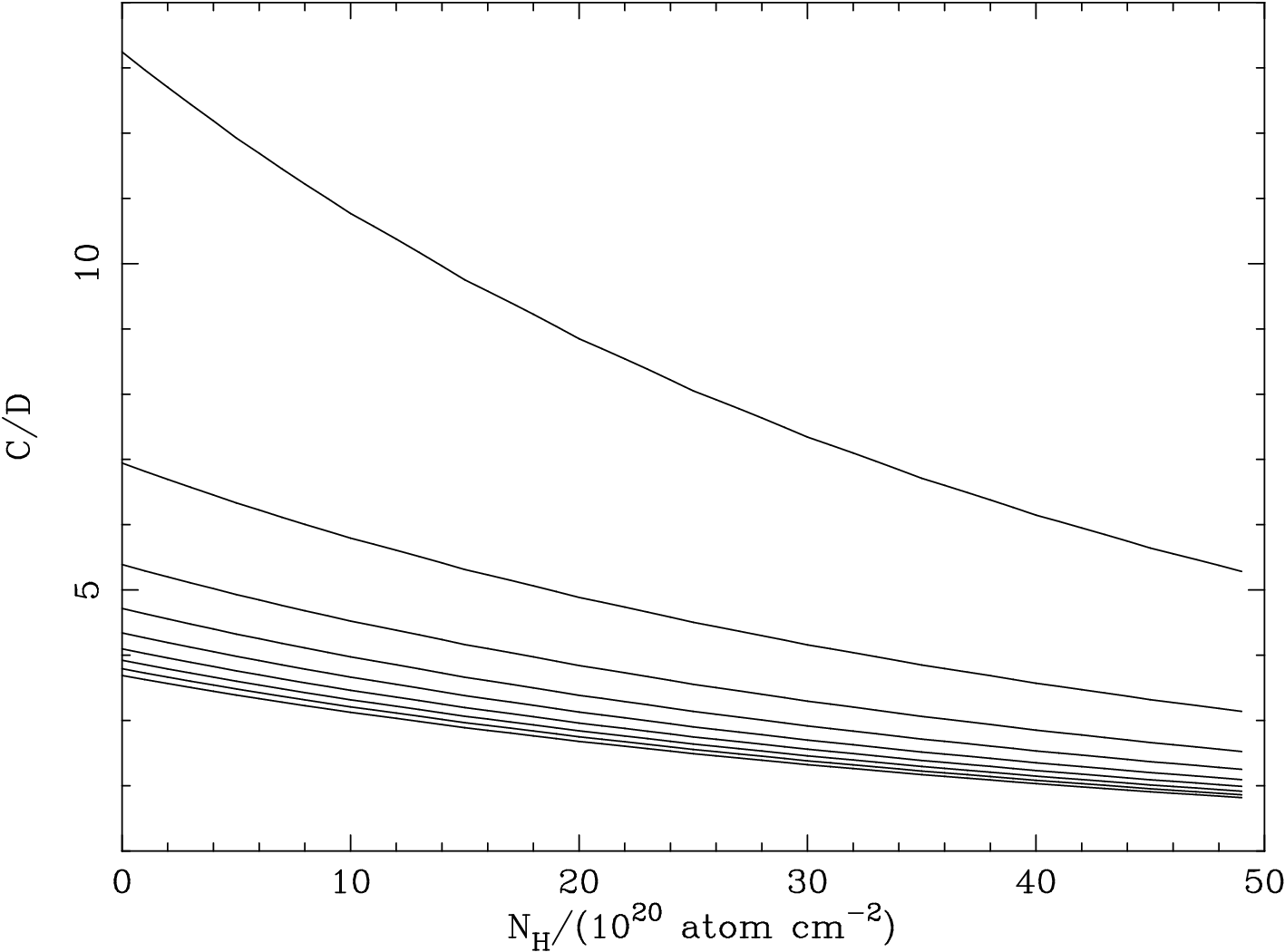}
  \end{tabular}
  \caption{Theoretical predictions for the $B/D$ and $C/D$ ratios for
  an gas cooling from temperatures of 1--10 keV to 0.001 keV. The lines show steps
  of 1 keV in temperature. The lowest temperature at the highest ratio
  on the $y$-axis. (a) $B/D$ and (b) $C/D$ as functions of metallicity
  at fixed column density ($N_H = 10^{20} \textrm{ atom
  cm}^{-2}$). (c) $B/D$ and (d) $C/D$ as functions of column density
  at fixed metallicity ($Z = 0.5 Z_{\odot}$).}
\label{fig:cfmodels}
\end{figure*}

\begin{figure*}
  \begin{tabular}{llll}
    (a) &
    \includegraphics[angle=-90,width=0.36\textwidth]{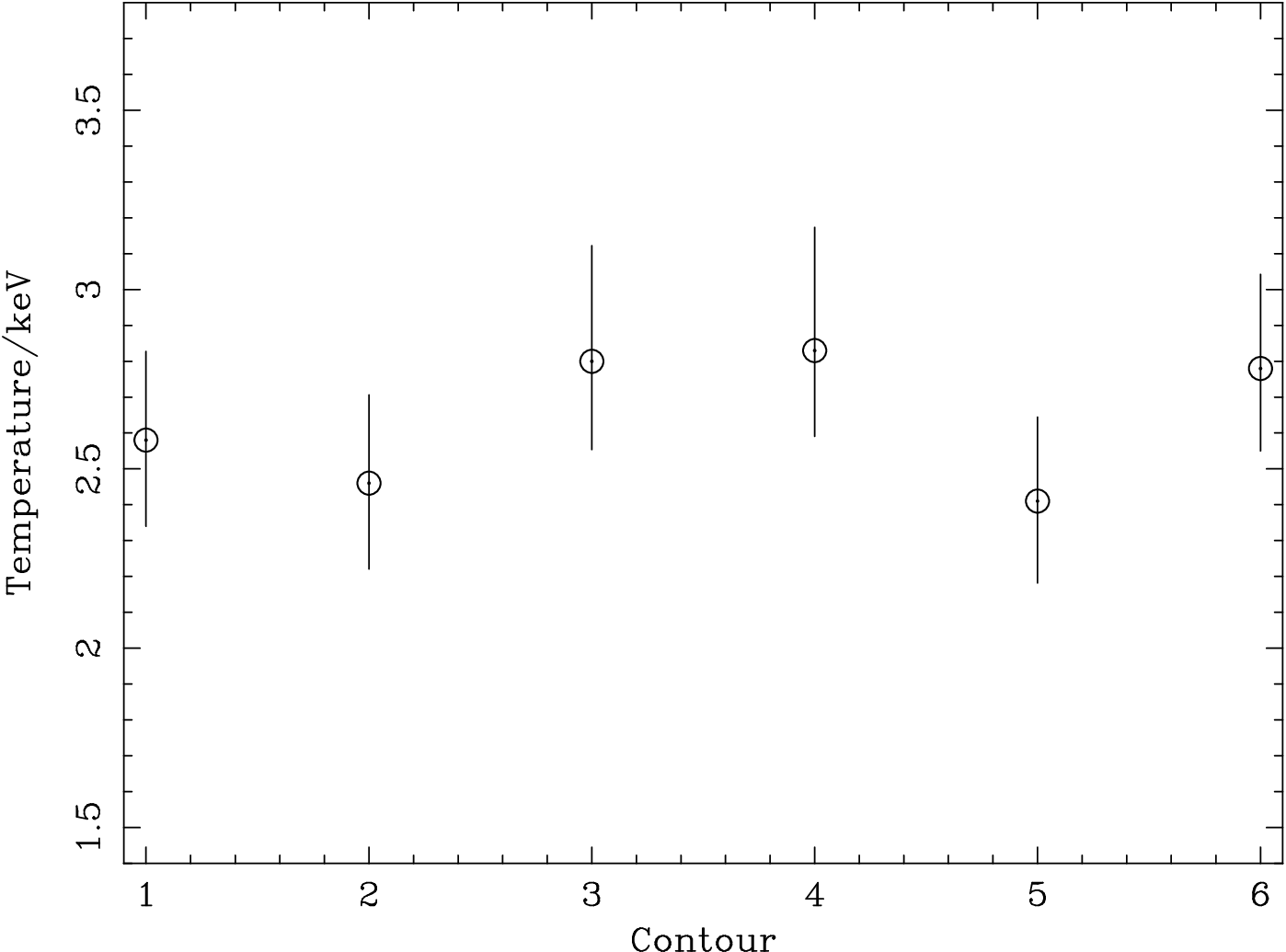} &
    (b) &
    \includegraphics[angle=-90,width=0.36\textwidth]{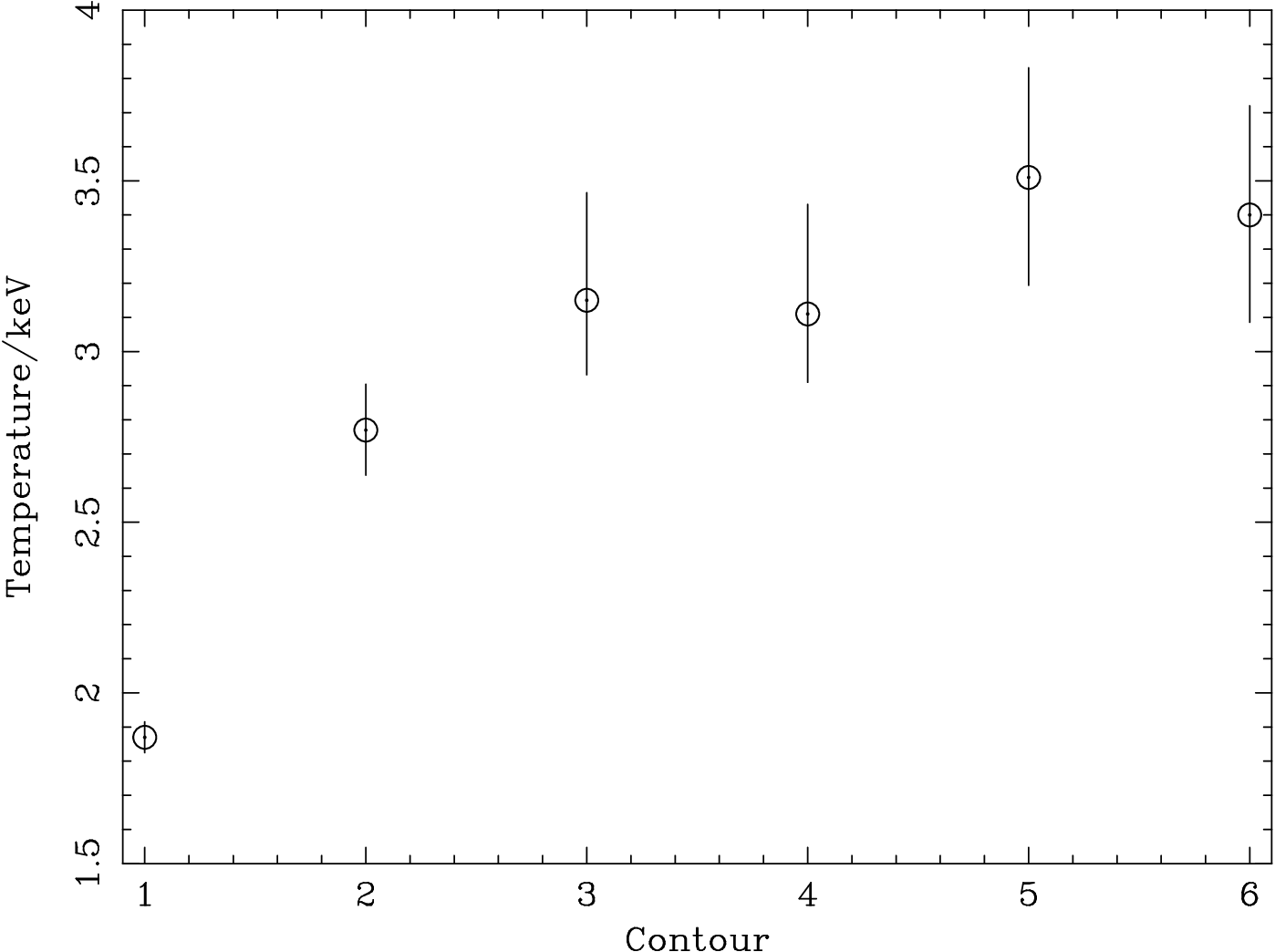} \\
    (c) &
    \includegraphics[angle=-90,width=0.36\textwidth]{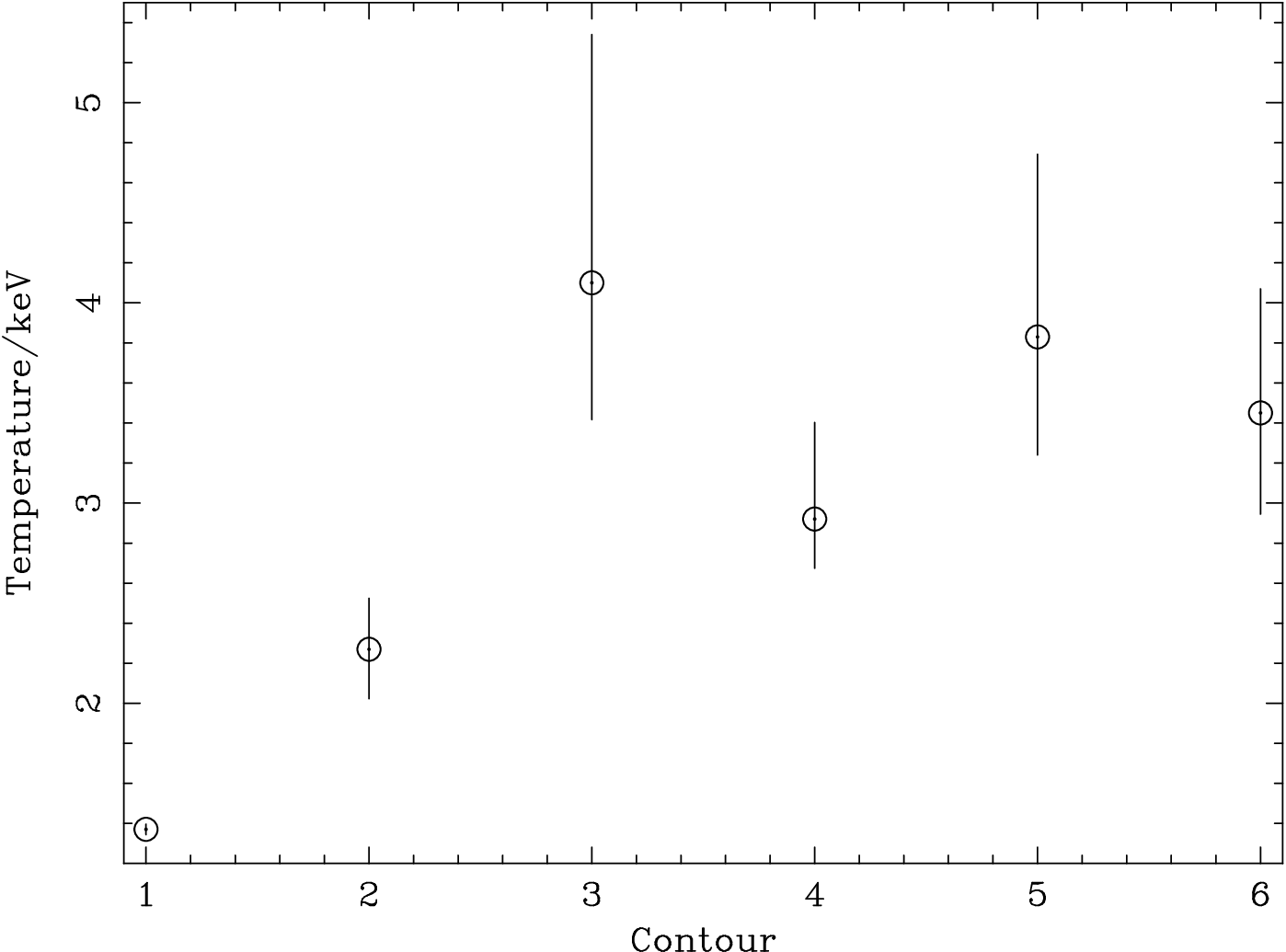} &
    (d) &
    \includegraphics[angle=-90,width=0.36\textwidth]{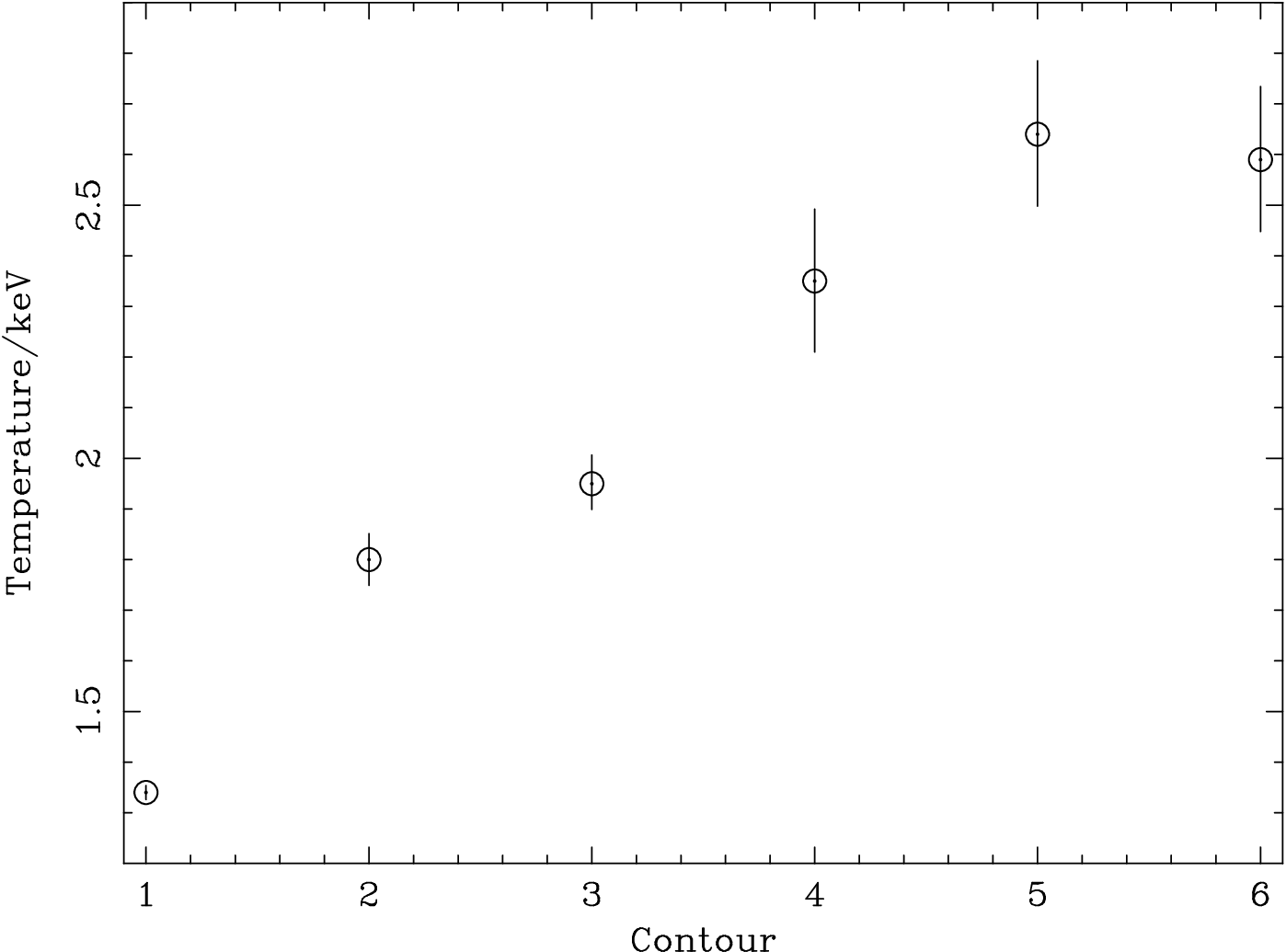}
  \end{tabular}
  \caption{Temperature profiles generated by fitting constant galactic
    obscuring column density and constant $0.3 \: Z_{\odot}$
    abundance to the $C/D$ colour profile. The clusters are (a) A1060
    (non-cooling-flow), (b) A2199, (c) Centaurus and (d) Virgo. The
    errors are 1-$\sigma$, generated from the errors in $C/D$. The
    contours are numbered from innermost to outermost.}
  \label{fig:tempprof}
\end{figure*}

\begin{figure*}
  \begin{tabular}{llll}
    (a) &
    \includegraphics[angle=-90,width=0.36\textwidth]{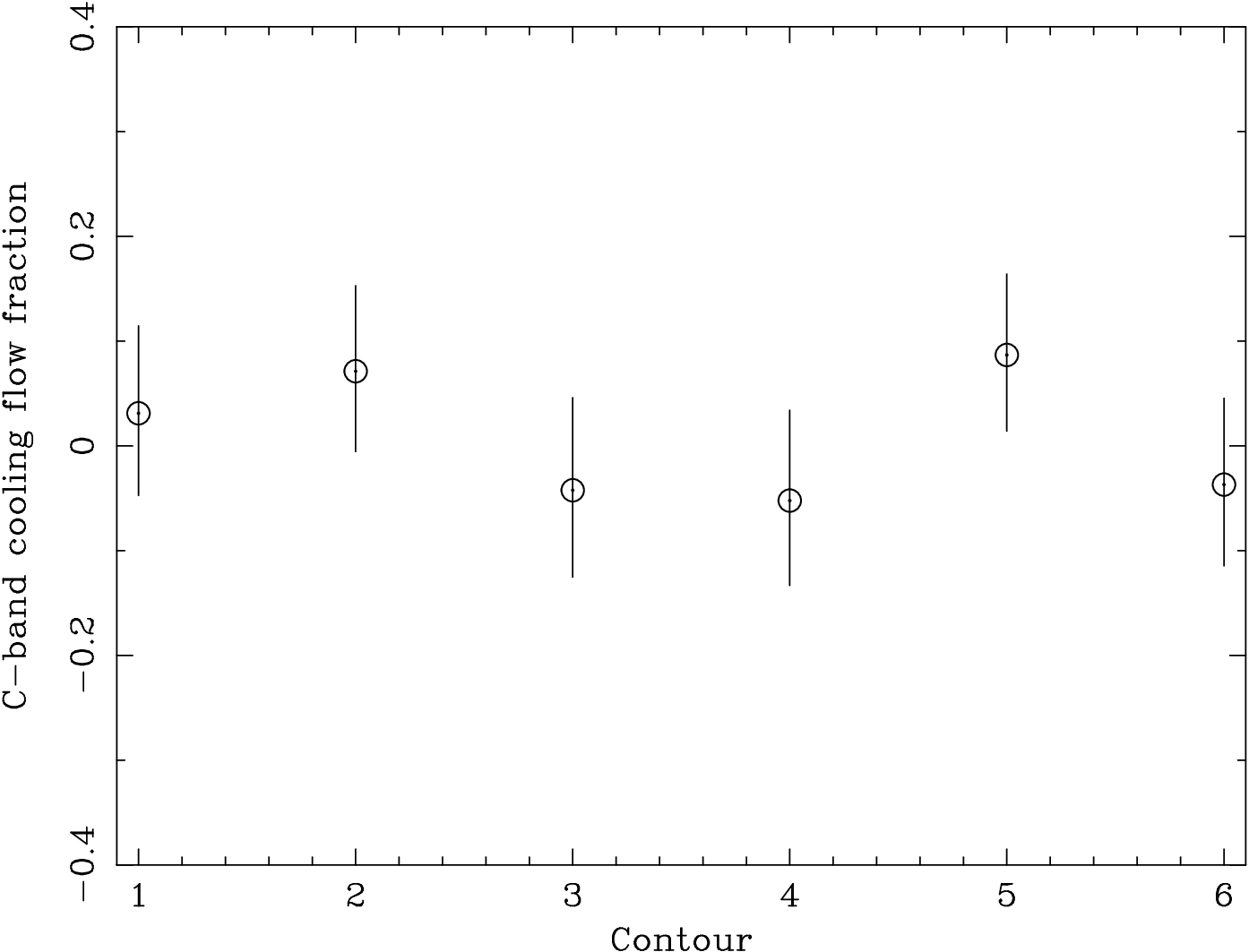} &
    (b) &
    \includegraphics[angle=-90,width=0.36\textwidth]{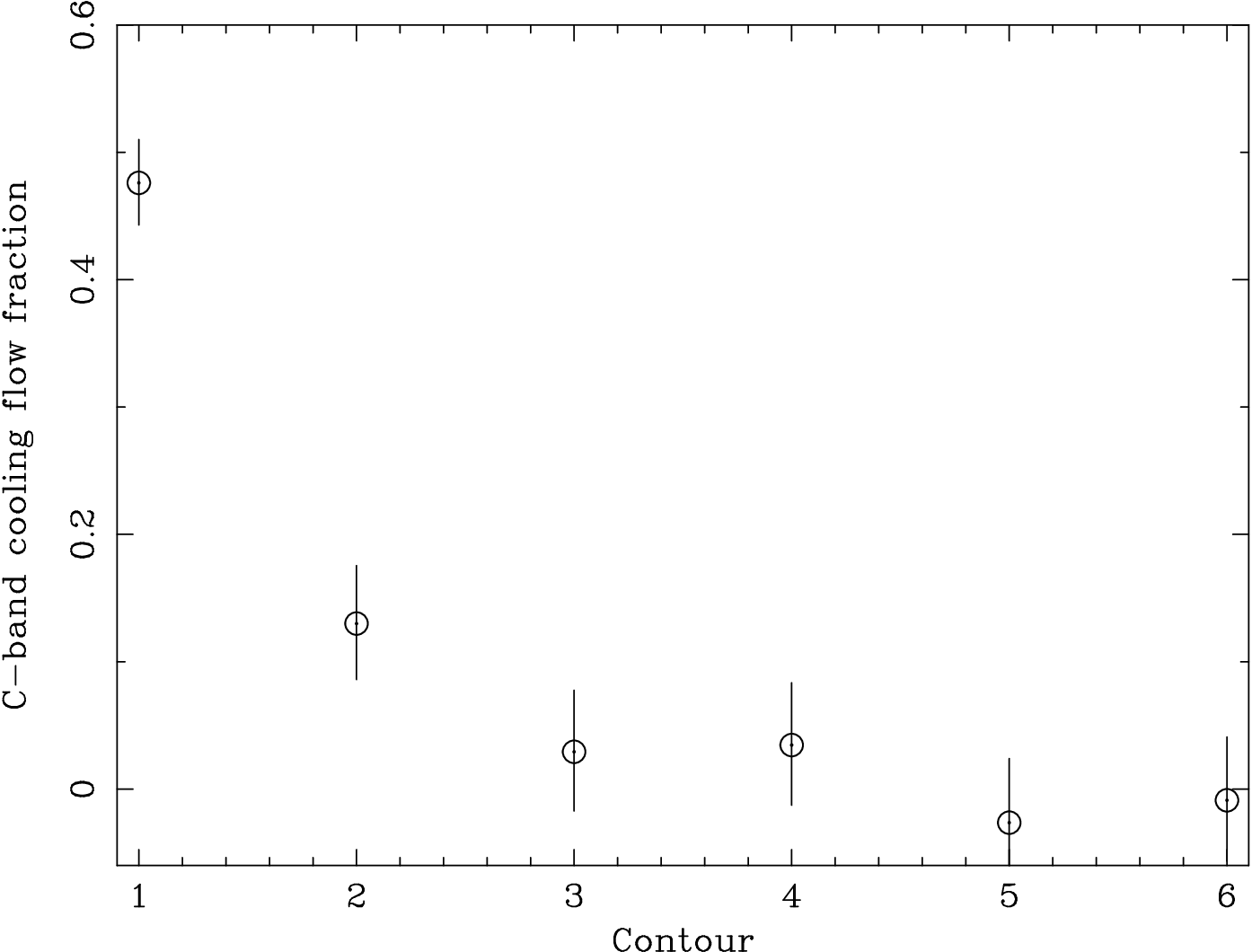} \\
    (c) &
    \includegraphics[angle=-90,width=0.36\textwidth]{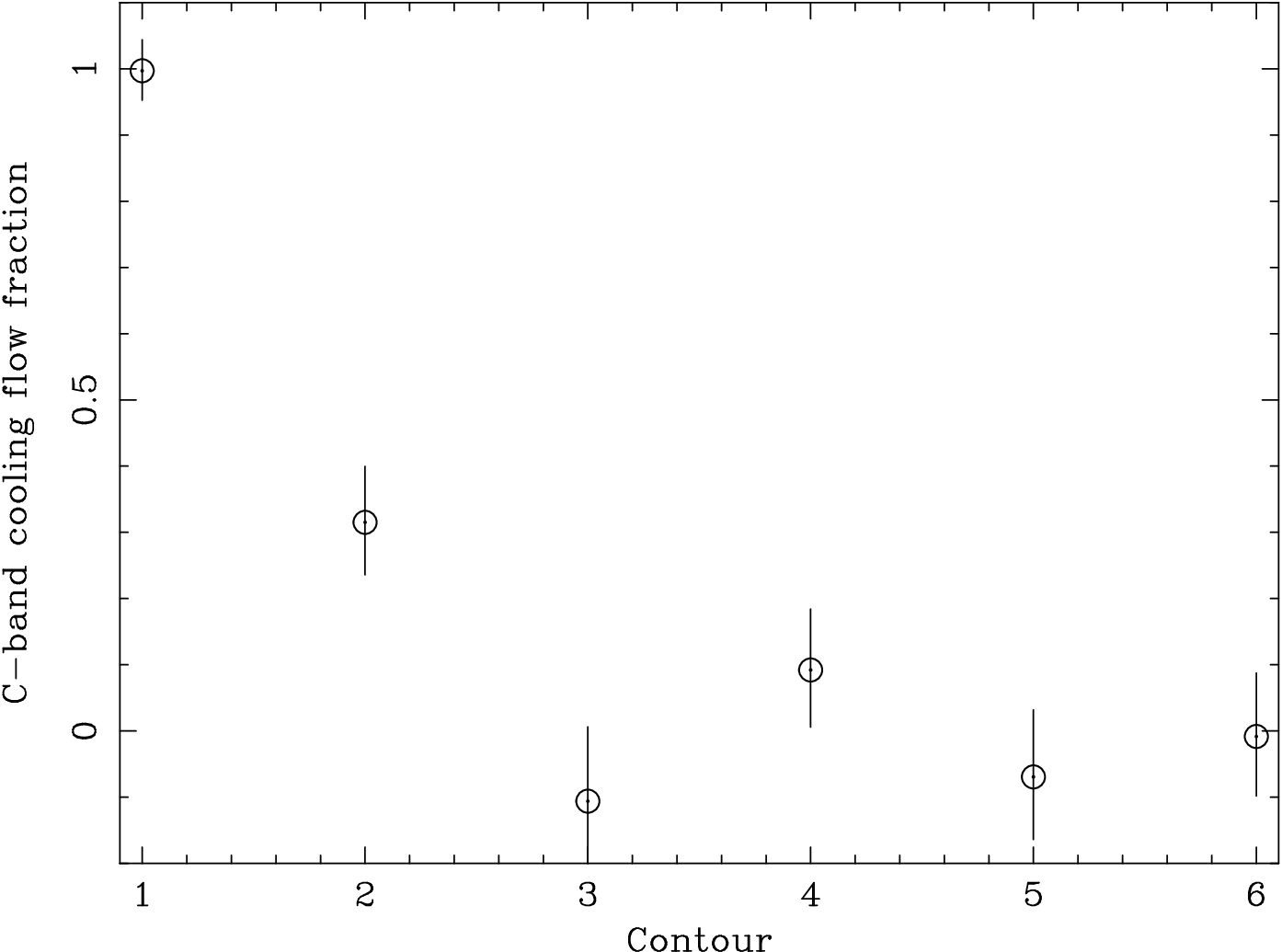} &
    (d) &
    \includegraphics[angle=-90,width=0.36\textwidth]{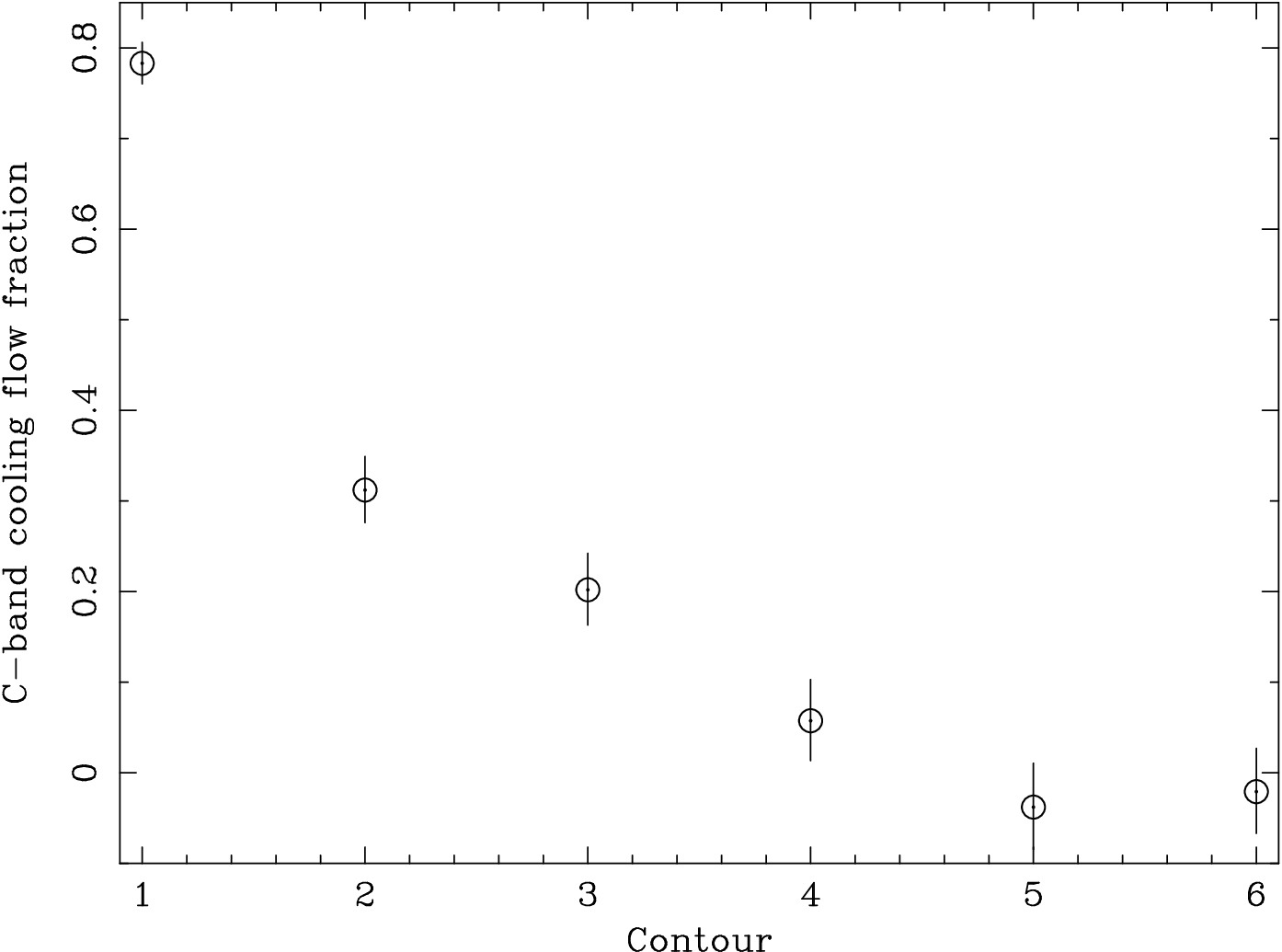}
  \end{tabular}
  \caption{Profiles showing how the fractional cooling flow component
    must increase to fit the $C/D$ ratio assuming an isothermal
    temperature, constant galactic column density and constant $0.3
    \: Z_{\odot}$ abundance. The clusters are (a) A1060
    (non-cooling-flow), (b) A2199, (c) Centaurus and (d) Virgo. The
    errors are 1-$\sigma$, generated from the errors in $C/D$. The
    contours are numbered from innermost to outermost.}
  \label{fig:cfprof}
\end{figure*}

\begin{figure*}
  \begin{tabular}{llll}
    (a) &
    \includegraphics[angle=-90,width=0.36\textwidth]{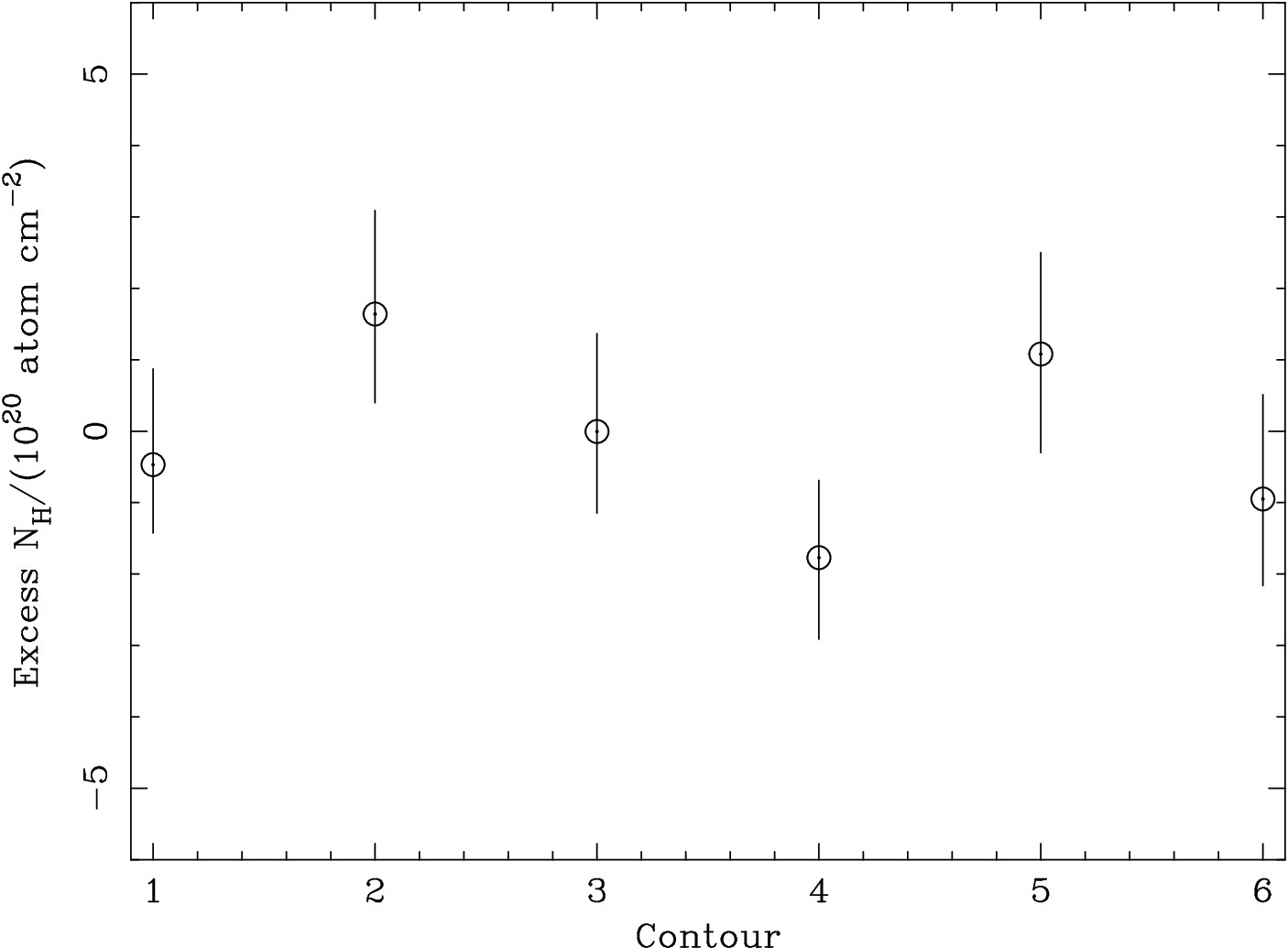} &
    (b) &
    \includegraphics[angle=-90,width=0.36\textwidth]{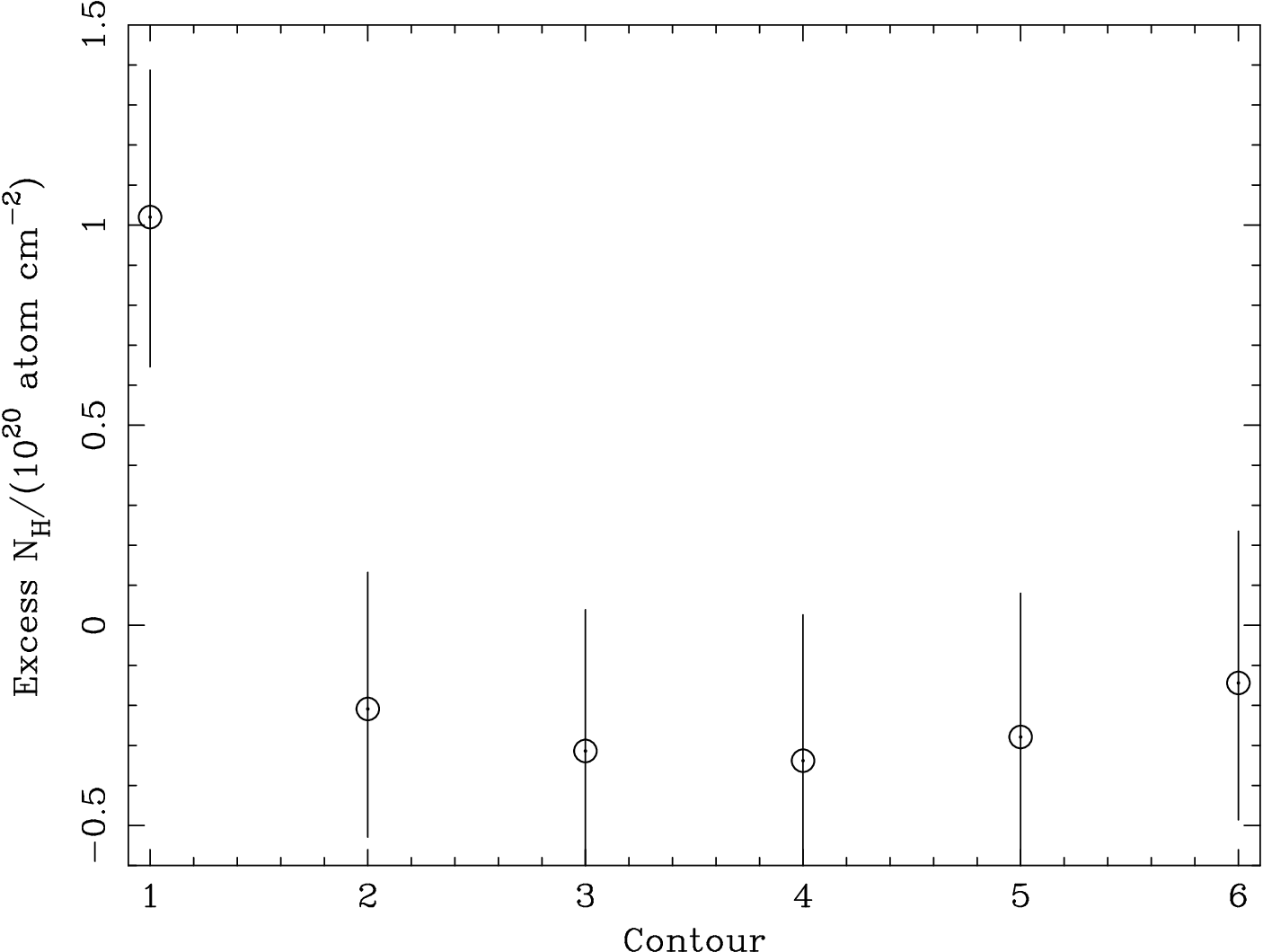} \\
    (c) &
    \includegraphics[angle=-90,width=0.36\textwidth]{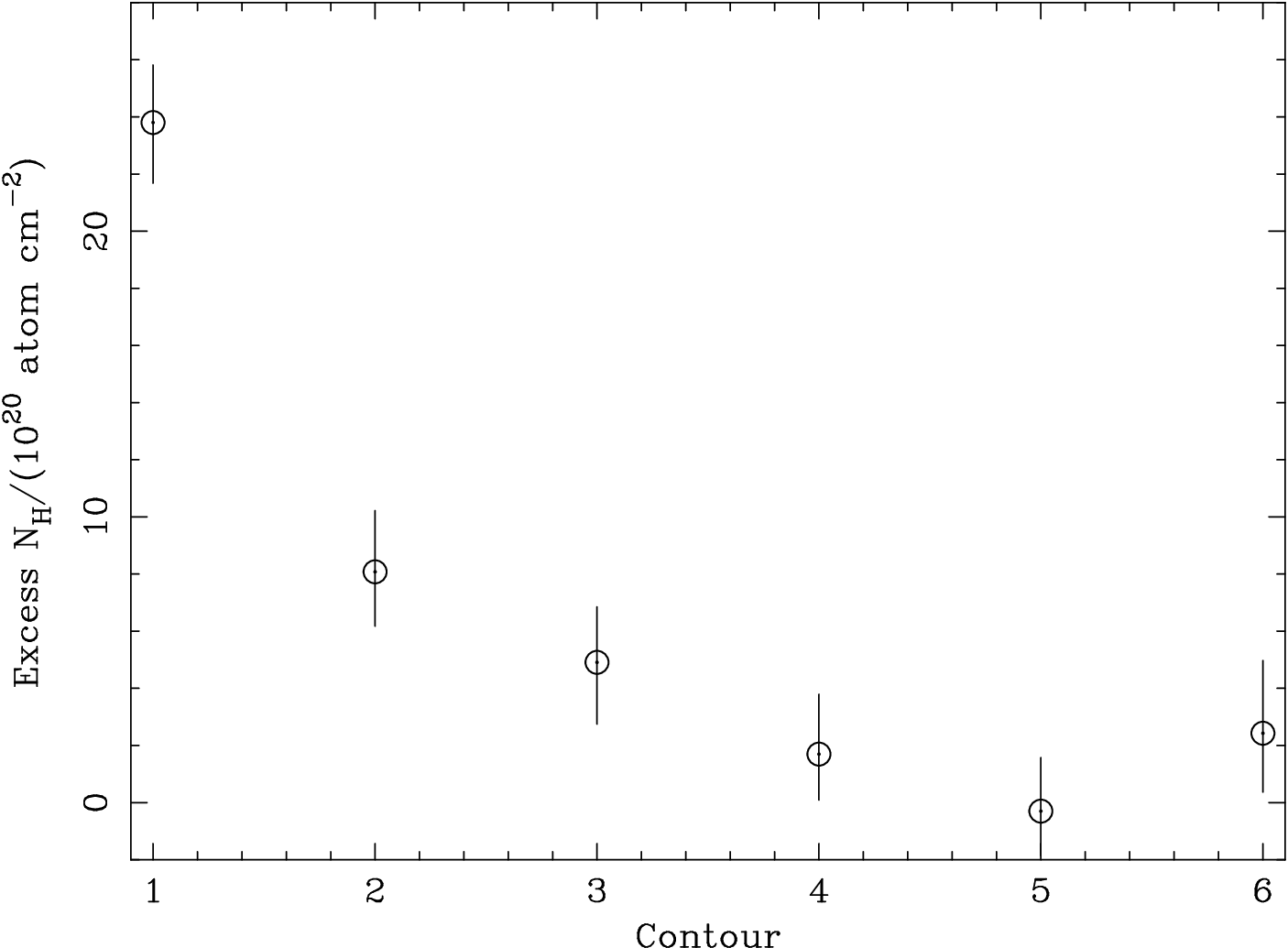} &
    (d) &
    \includegraphics[angle=-90,width=0.36\textwidth]{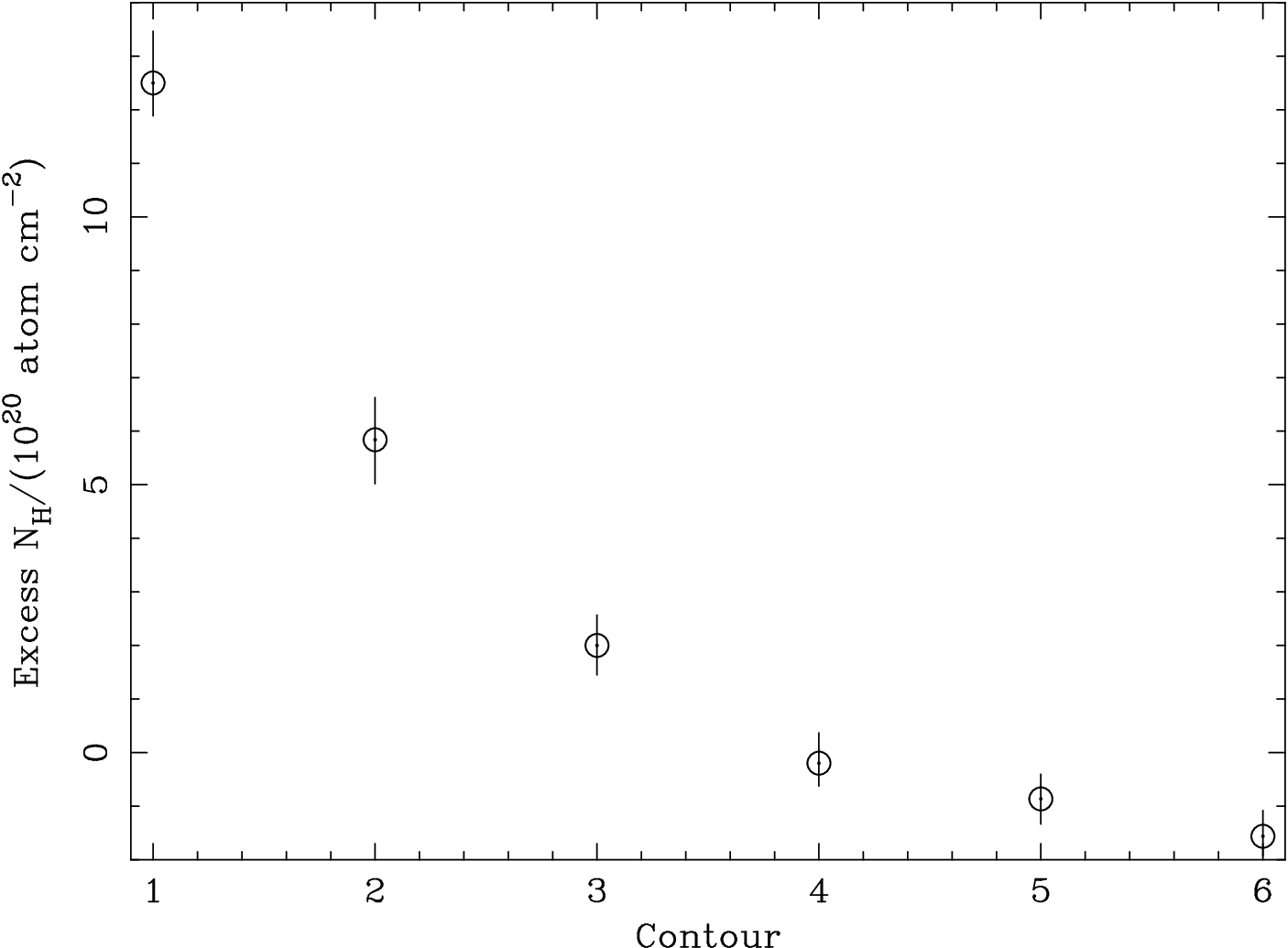}
  \end{tabular}
  \caption{Profiles showing how the intrinsic absorption must
    increase assuming an isothermal temperature, constant galactic
    column density and a constant $0.3 \: Z_{\odot}$ abundance.
    These were calculated by assuming the cooling flow
    components above, and calculating the extra required column density
    above galactic to explain the $B/D$ ratio, assuming constant $0.3
    \: Z_{\odot}$ abundance. The errors are 1-$\sigma$, and were
    generated from the error in $B/D$, the error in the cooling
    flow component is ignored. The clusters are (a) A1060
    (non-cooling-flow), (b) A2199, (c) Centaurus and (d) Virgo. The
    contours are numbered from innermost to outermost.} 
  \label{fig:deltanhprof}
\end{figure*}

\begin{table}
\caption{Table showing the clusters and their calculated excess
absorbtion, $\Delta N_H$ (units $10^{20} \: \mathrm{atom} \:
\mathrm{cm}^{-2}$). Also shown is the used galactic column density,
$N_{H,\mathrm{gal}}$
(units $10^{20} \: \mathrm{atom} \: \mathrm{cm}^{-2}$), and
the significance of the excess absorption, $S(\Delta N_H)$.
The clusters are listed in
cooling flux order ($\dot{M}/z^2$). The outer contour of A3571 was
ignored as it was suspected to have background
contamination. $T_{\mathrm{iso}}$ is the isothermal temperature
calculated from the outer contours, which is limited to a maximum
of 10 keV.}
\begin{tabular}{llrrr}
Cluster &$N_{H,\mathrm{gal}}$
                    & $\Delta N_H$        & $S(\Delta N_H)$ &
                                              $T_{\mathrm{iso}}$ / keV\\
\hline
Virgo      &   2.54 & $ 13.7\pm   0.7$ &  19.0  & $  2.5 \pm 0.1$ \\
Perseus    &   14.9 & $  3.1\pm   1.1$ &   2.8  & $  3.4 \pm 0.2$ \\
2A 0335+096 &  17.8 & $ 10.0\pm   4.5$ &   2.2  & $  3.1 \pm 0.4$ \\
Centaurus  &   8.06 & $ 22.7\pm   2.4$ &   9.3  & $  3.4 \pm 0.4$ \\
A2199      &   0.86 & $  1.2\pm   0.4$ &   2.9  & $  3.3 \pm 0.2$ \\
Ophiuchus  &   20.3 & $  4.7\pm   3.4$ &   1.4  & $ > 10.0$       \\
KLEM44     &   1.56 & $  4.1\pm   1.6$ &   2.5  & $  3.3 \pm 1.5$ \\
A2052      &   2.71 & $  4.7\pm   2.8$ &   1.7  & $  4.3 \pm 0.8$ \\
A262       &   5.37 & $ 13.1\pm   3.7$ &   3.6  & $  2.2 \pm 0.2$ \\
PKS 0745-191 & 35.0 & $ -5.0\pm   5.8$ &  -0.9  & $  3.2 \pm 0.7$ \\
A1060      &   5.47 & $ -0.5\pm   1.5$ &  -0.3  & $  2.7 \pm 0.2$ \\
A1795      &   1.19 & $  0.6\pm   0.4$ &   1.5  & $  5.8 \pm 0.8$ \\
Hydra A    &   4.94 & $  2.4\pm   1.4$ &   1.7  & $  4.1 \pm 0.9$ \\
A2029      &   3.05 & $  1.0\pm   1.3$ &   0.7  & $  7.0 \pm 1.1$ \\
A496       &   4.58 & $  5.3\pm   2.2$ &   2.4  & $  3.7 \pm 0.7$ \\
MKW3       &   3.04 & $  2.2\pm   2.2$ &   1.0  & $  3.8 \pm 1.3$ \\
A478       &   30.0 & $  9.0\pm   3.0$ &   3.0  & $  3.7 \pm 0.5$ \\
A85        &   3.45 & $  2.1\pm   1.5$ &   1.4  & $  4.8 \pm 1.3$ \\
A3112      &    4.0 & $  5.0\pm   1.7$ &   3.0  & $  3.8 \pm 1.9$ \\
Cygnus A   &   34.7 & $  5.4\pm   6.6$ &   0.8  & $  2.6 \pm 0.3$ \\
AWM7       &   9.81 & $  4.3\pm   1.6$ &   2.7  & $  3.0 \pm 0.2$ \\
A4059      &   1.10 & $  2.6\pm   2.4$ &   1.0  & $  6.1 \pm 1.6$ \\
A3571      &   3.71 & $ -3.8\pm   3.0$ &  -1.3  & $  7.7 \pm 0.2$ \\
A2597      &   2.49 & $  7.4\pm   2.9$ &   2.6  & $  3.3 \pm 1.7$ \\
A644       &   6.82 & $ -3.9\pm   2.9$ &  -1.3  & $  5.0 \pm 1.9$ \\
A2204      &   5.67 & $ -3.4\pm   5.3$ &  -0.6  & $  6.2 \pm 1.5$ \\
A2142      &   4.20 & $  0.7\pm   1.8$ &   0.4  & $  > 10.0 $     \\
A3558      &   3.89 & $  1.5\pm   1.0$ &   1.4  & $  3.5 \pm 0.3$ \\
A401       &   10.5 & $ -2.7\pm   3.3$ &  -0.8  & $  6.8 \pm 1.4$ \\
Coma       &  0.918 & $ -0.3\pm   0.4$ &  -0.8  & $  5.2 \pm 0.4$ \\
A754       &   4.37 & $  1.6\pm   1.8$ &   0.9  & $  7.4 \pm 0.7$ \\
A2256      &   4.10 & $  1.8\pm   3.2$ &   0.6  & $  7.7 \pm 0.8$ \\
A119       &   3.44 & $ -1.4\pm   1.6$ &  -0.8  & $  3.2 \pm 0.5$ \\
\hline
\end{tabular}
\label{tab:delnh}
\end{table}

\section{Colour models}
\label{sec:models}
By comparing the $B/D$ and $C/D$ ratios with
theoretical models, it is possible to estimate the effective
temperature, metallicity and absorbing column density. We reevaluated
the theoretical curves of Allen \& Fabian (1997) using a more recent
model of an isothermal gas; the \textsc{xspec 10} \textsc{mekal} model
based on the calculations of Mewe and Kaastra with Fe L calculations
by Liedahl (Mewe, Gronenschild \& van den Oord 1985; Liedahl,
Osterheld \& Goldstein 1995) with an absorbing \textsc{phabs} screen.

We iterated the calculation of the $B/D$ and $C/D$ colour ratios over
the parameter space $0 \leq N_H/(10^{20} \textrm { atom cm}^{-2}) \leq
20$, $0 \leq Z/Z_{\odot} \leq 2$, and $1 \leq T/\textrm{keV} \leq 10$,
where $N_H$ is the column density of the absorbing screen, and $Z$
and $T$ are the metallicity and temperature of the cluster.
Projections of the ratios for constant $Z$ or $N_H$ are shown in
Fig. \ref{fig:isomodels}. The colours were calculated for an object of
redshift of 0.01. We found that the choice of redshift was not
significant, due to the low redshift nature of the clusters examined.
The predictions of the model show some differences from those
of Allen \& Fabian (1997), particularly at low temperatures, where the
colour ratios are higher than those predicted before.

We calculated, too, the colour ratios expected for a cooling flow
model (Johnstone et al. 1992) with an absorbing screen
(Fig. \ref{fig:cfmodels}). We modelled a gas cooling to 0.001 keV over
the same range of abundance, upper temperature and absorbing column
density as the isothermal gas model.

Our first step was to fit the $C/D$ colour ratios for the contours of 
a cluster with
a single component isothermal model. We assumed galactic absorption,
$N_{H,\mathrm{gal}}$, for each cluster and a constant metallicity of
$Z_{\mathrm{clust}}=0.3 \: Z_{\odot}$. By taking the average fitted
temperature of the outer three contours we derived an isothermal
temperature of each cluster, $T_{\mathrm{iso}}$. Table
\ref{tab:delnh} shows the fitted isothermal temperature for each
cluster, with the galactic absorption used. The galactic column
densities listed were obtained using the \textsc{ftools} \textsc{nh}
program (Dickey and Lockman 1990), except for A478, PKS 0745-191 and
A3112, for which we used the values quoted in White (2000).

Temperature profiles calculated using an isothermal model, with fixed
absorption and metallicities, are shown for a sample of four clusters
in Fig. \ref{fig:tempprof}.  The errors in the temperatures were
propagated directly from the errors in the observed $C/D$ ratio only.
Three of the clusters are known cooling flow clusters. Applying the
isothermal model to each contour shows a temperature decrease from the
outer regions of the clusters to their centres.  For the
non-cooling-flow cluster there is no evidence of this temperature
change. This is generally the case for our sample of clusters. Those
with strong cooling flows show a temperature decrease, except for
those masked by a strong galactic column.

\subsection{Fitting metallicity gradients}
The cooling flow clusters show a temperature decrease from their outer
regions to their centres, assuming a single-phase isothermal
gas. White (2000) found that 90 per cent of his sample of 98 clusters
were consistent with isothermality (after including cooling flow
effects) at the 3-$\sigma$ level. We attempted to fit the contour
colours with an isothermal gas with a varying metallicity. We
minimized the function
\begin{eqnarray}
  \chi^2(Z) & = & 
  \left\{
      \frac{ [B/D]_{\mathrm{obs}} - [B/D]_{\mathrm{iso}}
        (Z, T_{\mathrm{iso}}, N_{H,\mathrm{gal}}) }
      {\sigma_{ [B/D]_{\mathrm{obs}}}}
      \right\}^2 \nonumber \\
    & + &
  \left\{
      \frac{ [C/D]_{\mathrm{obs}} - [C/D]_{\mathrm{iso}}
        (Z, T_{\mathrm{iso}}, N_{H,\mathrm{gal}}) }
      {\sigma_{ [C/D]_{\mathrm{obs}}}}
      \right\}^2
\end{eqnarray}
to find the optimum value of $Z$, where $[B/D]_{\mathrm{obs}}$ and
$[C/D]_{\mathrm{obs}}$ are the observer contour colour values,
$\sigma_{ [B/D]_{\mathrm{obs}}}$ and $\sigma_{ [C/D]_{\mathrm{obs}}}$
are their respective errors, and $[B/D]_{\mathrm{iso}}$ and
$[C/D]_{\mathrm{iso}}$ are the predicted isothermal values as a function
of $Z$.

This analysis showed that simply using an isothermal gas with a
metallicity gradient was not a good fit to the observed
data in the inner parts of the clusters. The fitted metallicity was
$0.9Z_{\odot}$ for the
inner contour of the Virgo cluster data,
but the $\chi^2$ value was $\sim 600$. Most
of the other strong cooling flows showed similar results.

\subsection{Fitting cooling flow models}
To fit our data it was clear that another component to the model was
required. We added an extra cooling flow component, which was an
obvious first candidate. If the number of counts
observed in a particular band is denoted by $X_{\mathrm{obs}}$, an
isothermal model predicts a count of $X_{\mathrm{iso}}(Z, T,
N_H)$, a cooling flow model predicts $X_{\mathrm{cf}}(Z, T, N_H)$,
where $T$ in this case denotes the upper cooling temperature, and
the models are normalised by the values $N_{\mathrm{iso}}$ and
$N_{\mathrm{cf}}$, then we can write
\begin{eqnarray}
  B_{\mathrm{obs}} &=& N_{\mathrm{iso}} \: B_{\mathrm{iso}}(Z, T, N_H) +
   N_{\mathrm{cf}} \: B_{\mathrm{cf}}(Z, T, N_H),
  \label{eqn:bcounts} \\
  C_{\mathrm{obs}} &=& N_{\mathrm{iso}} \: C_{\mathrm{iso}}(Z, T, N_H) +
   N_{\mathrm{cf}} \: C_{\mathrm{cf}}(Z, T, N_H), \textrm{ and}
  \label{eqn:ccounts} \\
  D_{\mathrm{obs}} &=& N_{\mathrm{iso}} \: D_{\mathrm{iso}}(Z, T, N_H) +
   N_{\mathrm{cf}} \: D_{\mathrm{cf}}(Z, T, N_H).
  \label{eqn:dcounts}
\end{eqnarray}

By solving the above equations we can find the cooling flow fraction in a
particular band. In Band $C$ the cooling flow fraction is
\begin{eqnarray}
 f_{C} & = & \frac{N_{\mathrm{cf}} \: C_{\mathrm{cf}}}
 { N_{\mathrm{iso}} \: C_{\mathrm{iso}} + 
   N_{\mathrm{cf}} \: C_{\mathrm{cf}}}
 \nonumber \\
 & = & \frac{  C_{\mathrm{cf}} ( C_{\mathrm{iso}} -
 \left[ \frac{C}{D} \right]_{\mathrm{obs}} \:  D_{\mathrm{iso}} )}
  {\left[ \frac{C}{D} \right]_{\mathrm{obs}} \: (
     D_{\mathrm{cf}} \:  C_{\mathrm{iso}} -
     C_{\mathrm{cf}} \:  D_{\mathrm{iso}})
     }.
\end{eqnarray}
Note that the model predictions, $X_{\mathrm{iso}}$ and
$X_{\mathrm{cf}}$ are functions of metallicity, absorbing column
density and temperature. We made the assumption of constant
metallicity, isothermal temperature (or upper cooling temperature)
and obscuration by galactic column.
Those clusters with strong cooling flows show large cooling flow
fractions at their centres. In Fig. \ref{fig:cfprof} we present
$C$ band cooling flow fraction profiles for our four example
clusters. The errors shown only take account of the uncertainties in the
$C/D$ ratio.

\subsection{Calculating intrinsic absorption}
By rearranging equations (\ref{eqn:bcounts}), (\ref{eqn:ccounts}) and
(\ref{eqn:dcounts}), we can predict the observed $B/D$ colour from the
$C/D$ colour and the models. Rearranging,
\begin{equation}
  \left[ \frac{B}{D} \right]_{\mathrm{pred}} =
  \frac{
    \left[ \frac{C}{D} \right]_{\mathrm{obs}} -
    \frac{ C_{\mathrm{cf}} }{ D_{\mathrm{cf}} }
    }{
    \frac{ C_{\mathrm{iso}} } { B_{\mathrm{iso}}} -
    \frac{ C_{\mathrm{cf}} } { B_{\mathrm{iso}}}
    \frac{ D_{\mathrm{iso}} } { D_{\mathrm{cf}}}
    }
  +
  \frac{
    \left[ \frac{C}{D} \right]_{\mathrm{obs}} -
    \frac{ C_{\mathrm{iso}} }{ D_{\mathrm{iso}} }
    }{
    \frac{ C_{\mathrm{cf}} } { B_{\mathrm{cf}}} -
    \frac{ C_{\mathrm{iso}} } { B_{\mathrm{cf}}}
    \frac{ D_{\mathrm{cf}} } { D_{\mathrm{iso}}}
    }.
\label{eqn:predbd}
\end{equation}

We found that the predicted $B/D$ colour does not match the observed
value. The $B/D$ ratio is more sensitive to absorption than the $C/D$
ratio (Fig. \ref{fig:isomodels} and \ref{fig:cfmodels}). Assuming the
difference between the observed and predicted values of $B/D$ is due
to intrinsic absorption, then it is possible to use the above equation
to find the increase in absorption. We varied the obscuring column
density to find $\{ B,C,D \}_{ \{ \mathrm{iso},\mathrm{cf} \} }$ until
the predicted $B/D$ ratio from equation (\ref{eqn:predbd}) was the
same as the observed one.

We then plotted a profile showing the required extra absorption above
galactic as a function of contour for each cluster.  Those clusters
with strong cooling flows show an increase in absorption at their
centres. Fig.  \ref{fig:deltanhprof} shows profiles of the required
extra absorption for our example four clusters. Listed in Table
\ref{tab:delnh} are the required absorption increases for each cluster
in our sample. This value is the difference between the absorption in
the central contour and the mean of the two outer ones. We took the
mean outer contours to decrease the chance of background
contamination. We looked at the difference in absorption  between the
centre and outside of the cluster, rather than the absolute difference
from the central absorption to galactic, because the galactic
absorption value is uncertain for many clusters. The errors on the
values assume that all uncertainty is due to the observed $B/D$
ratio. $S(\Delta N_H)$ shows the significance of the increased
absorption using an expression with the same form as equation
(\ref{eqn:sig}).

\section{Discussion}
\subsection{Adaptively-smoothed images}
Many of the cooling flow clusters in our sample show relatively featureless
elliptical images after adaptive smoothing. We present some interesting
ones, and those of the Perseus, Virgo and A2199 clusters in Fig.
\ref{fig:per_adapt}, \ref{fig:vir_adapt} and \ref{fig:a2199_adapt}.
The clusters show structure which appears to vary between the different
bands.

\subsection{The sample of clusters}
Fig. \ref{fig:b_d_corr} and \ref{fig:c_d_corr} show that clusters with
low galactic latitude, or those masked by regions of high absorption,
show low $B/D$ and $C/D$ colours. These clusters include, for example,
Ophiuchus, PKS 0745-191 and Cygnus A.

Those clusters with strong cooling flows show large $C/D$ ratios, or
softer X-rays, at their centres, with large $C/D$ gradients moving
towards their centres. These include, for example, Virgo, Perseus, 2A
0335+096, Centaurus and A2199. They therefore contain cool gas at
small radii, within a few arcminutes of the centre.  This can be seen
by comparison with the theoretical curves in Fig.
\ref{fig:isomodels}. The only way of achieving the gradient is by
adding cooler gas. Those clusters without strong cooling flows do not
show this strong colour gradient, for example, Coma, A2256 and A119.

\subsection{Comparison with models}
The isothermal temperatures derived using the isothermal model in the
outer regions of the clusters agree in most cases with those derived
from $ASCA$ data (White 2000). There are a few clusters, most of which
are absorbed by high galactic column densities, for which we find
unusually low temperatures. These include Perseus, PKS 0745-191, A478
and Cygnus-A. We note that we were able to reproduce the Perseus $ASCA$
temperature result using a column of $10^{21}$ atom $\textrm{cm}^{-2}$.

Those clusters with strong cooling flows show good evidence of
absorbing material at their centres. For example Virgo, Centaurus,
Perseus and A478 all show significant increased levels of
absorption. There is no evidence for absorbing material in those
clusters without cooling flows.  Abundance gradients in cooling
flow clusters have been observed, for example in the Centaurus cluster
(Fukazawa et al. 1994), the Virgo cluster (Matsumoto et al. 1996) and
AWM7 (Ezawa et al. 1997), with the metallicities decreasing outwards
from the cluster centres. Due to the relationship on $B/D$ colour from
absorption and metallicity, there is some degeneracy in the two
variables. At low temperatures metallicity has its strongest effect
(Fig. \ref{fig:isomodels}), so if we have overestimated the levels of
absorption, then it will be primarily for those clusters with metallicity
gradients. However, we found that metallicity gradients were not
sufficient by themselves to account for the observed colour gradients.

The PSPC detector limits the energy bounds of our observations. There
are too few independent energy bands to make more than two
colours, which limits the number of physical quantities we can fit the
data. More colours will help to completely resolve any
degeneracy between metallicity and absorbing column
density. Data from \emph{Chandra} will be able to solve this
completely. The increased effective area of the telescope will also
improve our statistics, reducing the uncertainties in our colours.

Allen \& Fabian (1997) used a partial screening model in conjunction
with a deprojection method on their data. Using a partial screening
model results in larger column densities than those predicted without.
Doing a similar analysis would show larger levels of absorption in our
clusters, however it would add unnecessary complexity to our
procedure. We also did not take the absorption results to check for
consistency with the $C/D$ ratio, but do not expect the results to be
significantly different. 

\section{Conclusions}
We have presented an analysis of X-ray colour maps. The analysis
technique attempts to group areas of the image together in order to
maximize the signal to noise of the results, but preserving
information about the structure of the object. This technique will be
useful in examining data from the current generation of new X-ray
telescopes. These data will show large variations in count rate, and
will need analysis which is more intelligent than simple binning.

We created colour X-ray maps of the cores of a sample of 33 clusters,
almost doubling the sample of Allen \& Fabian (1997). The profiles
generated from these maps show that there is cooling gas at the
centres of those clusters with strong cooling flows. There is no
evidence for cooling gas in clusters inferred from imaging
deprojection or other methods to have weak, or no, cooling flow.  We
find a clear anti-correlation between the central cooling time of a
cluster and the $C/D$ colour gradient significance. The central
cooling time is an indicator of the quality of cooling flow data.  We
also find a correlation between the cooling flow flux and the $C/D$
significance.  The maps of the non-cooling-flow clusters contain more
structure as a group than those clusters with cooling flows.

We fitted a single-phase isothermal model to the $C/D$ colour of each
of the contours of our sample of clusters. Those clusters with strong
cooling flows show a significant decrease in the fitted temperatures
of their inner contours relative to their outer contours. An
isothermal gas with a varying metallicity alone was also not able to
fit our data. We added a cooling flow component to the model with the
same upper temperature as the isothermal plasma. Those clusters with
cooling flows required the addition of a significant fraction of
cooling flow model in their centres to account for the $C/D$
colour. To fit the observed $B/D$ colour we required the addition of
extra absorbing material at the centres of our strong cooling flow
clusters. This provides more evidence that cooling flows accumulate
cold material. There was no evidence for intrinsic absorbing material
for clusters without cooling flows.

\section*{Acknowledgements}
ACF and JSS thank the Royal Society and PPARC for support,
respectively. This research made use of the LEDAS archive of Leicester
University. The authors would like to thank the referee for his
constructive comments on the original manuscript.

\end{document}